\documentclass[10pt,twocolumn,letterpaper]{article}
\usepackage{amsmath,amsfonts}
\usepackage{algorithmic}
\usepackage{algorithm}
\usepackage{array}
\usepackage[caption=false,font=normalsize,labelfont=sf,textfont=sf]{subfig}
\usepackage[accsupp]{axessibility} 
\usepackage{ulem} 
\usepackage{textcomp}
\usepackage{epsfig}
\usepackage{stfloats}
\usepackage{verbatim}
\usepackage{booktabs}
\usepackage{graphicx}
\usepackage{cite}
\usepackage{color}
\usepackage{booktabs}
\usepackage{pifont}
\usepackage{bbding}
\usepackage{amssymb}
\usepackage{makecell}
\usepackage{amsthm,amsmath,amssymb}
\usepackage{mathrsfs}
\usepackage{threeparttable}
\usepackage{multirow,booktabs}
\usepackage{makecell}
\usepackage{multirow}
\newcommand\blue[1]{\textcolor{blue}{#1}}

\usepackage{cvpr}              % To produce the CAMERA-READY version
\usepackage[dvipsnames]{xcolor}
\newcommand{\red}[1]{{\color{red}#1}}

\usepackage{ulem}
\newcommand{\R}[1]{\textcolor[rgb]{1.00,0.00,0.00}{#1}}
\newcommand{\B}[1]{\textcolor[rgb]{0.00,0.00,1.00}{#1}}

\usepackage{makecell}
\usepackage{graphicx}
\usepackage{comment}

\definecolor{cvprblue}{rgb}{0.21,0.49,0.74}
\usepackage[pagebackref,breaklinks,colorlinks,citecolor=cvprblue]{hyperref}
\usepackage{indentfirst}

\def\paperID{14258} % *** Enter the Paper ID here
\def\confName{CVPR}
\def\confYear{2024}

\title{CPGA: Coding Priors-Guided Aggregation Network for Compressed Video Quality Enhancement}

\author{Qiang Zhu$^{1}$, Jinhua Hao$^{2}$, Yukang Ding$^{2}$, Yu Liu$^{1}$, Qiao Mo$^{1}$, Ming Sun$^{2}$, Chao Zhou$^{2}$, Shuyuan Zhu$^{1,}$\thanks{Corresponding author}\\
$^{1}$University of Electronic Science and Technology of China\\
$^{2}$Kuaishou Technology\\
{\tt\small \{zhuqiang@std.,202211012315@std.,eezsy@\}uestc.edu.cn, mqiao568@gmail.com}
\and
{\tt\small {\{haojinhua,dingyukang,sunming03,zhouchao\}@kuaishou.com}
}}

\begin{document}
\maketitle
\begin{abstract}

Recently, numerous approaches have achieved notable success in compressed video quality enhancement (VQE).
However, these methods usually ignore the utilization of valuable coding priors inherently embedded in compressed videos, such as motion vectors and residual frames, which carry abundant temporal and spatial information. 
To remedy this problem, we propose the \textbf{C}oding \textbf{P}riors-\textbf{G}uided \textbf{A}ggregation (CPGA) network to utilize temporal and spatial information from coding priors. The CPGA mainly consists of an inter-frame temporal aggregation (ITA) module and a multi-scale non-local aggregation (MNA) module. Specifically, the ITA module aggregates temporal information from consecutive frames and coding priors, while the MNA module globally captures spatial information guided by residual frames.
In addition, to facilitate research in VQE task, we newly construct the \textbf{V}ideo \textbf{C}oding \textbf{P}riors (VCP) dataset, comprising 300 videos with various coding priors extracted from corresponding bitstreams. It remedies the shortage of previous datasets on the lack of coding information. Experimental results demonstrate the superiority of our method compared to existing state-of-the-art methods. The code and dataset will be released at \url{https://github.com/VQE-CPGA/CPGA.git}.

\end{abstract}
    
\section{Introduction}
\label{sec:intro}
\vspace{-0.5em}

Video contains and communicates perceptual information derived from the real world. With the growth of the Internet, it has witnessed explosive growth in video content in digital network traffic~\cite{ref53}.
When transmitting videos over the Internet with limited bandwidth, efficient video codecs, such as H.264/AVC~\cite{ref52}, H.265/HEVC~\cite{ref1}, AV1~\cite{av1} and H.266/VVC~\cite{ref2,ref64} are widely used to save the coding bitrate.
However, due to the quantization in the lossy video coding, compressed videos inevitably exhibit compression artifacts, significantly diminishing the quality of experience~\cite{ref3,ref4,ref5,ref60,ref61,ref62,ref63}.
Moreover, compression artifacts may degrade the accuracy of vision tasks applied to videos, including object detection~\cite{ref6, ref69} and action recognition~\cite{ref7, ref70}.
Hence, improving the quality of compressed videos is of importance and practical significance.

\begin{figure}[!t]
\centering
\includegraphics[width=3.3in]{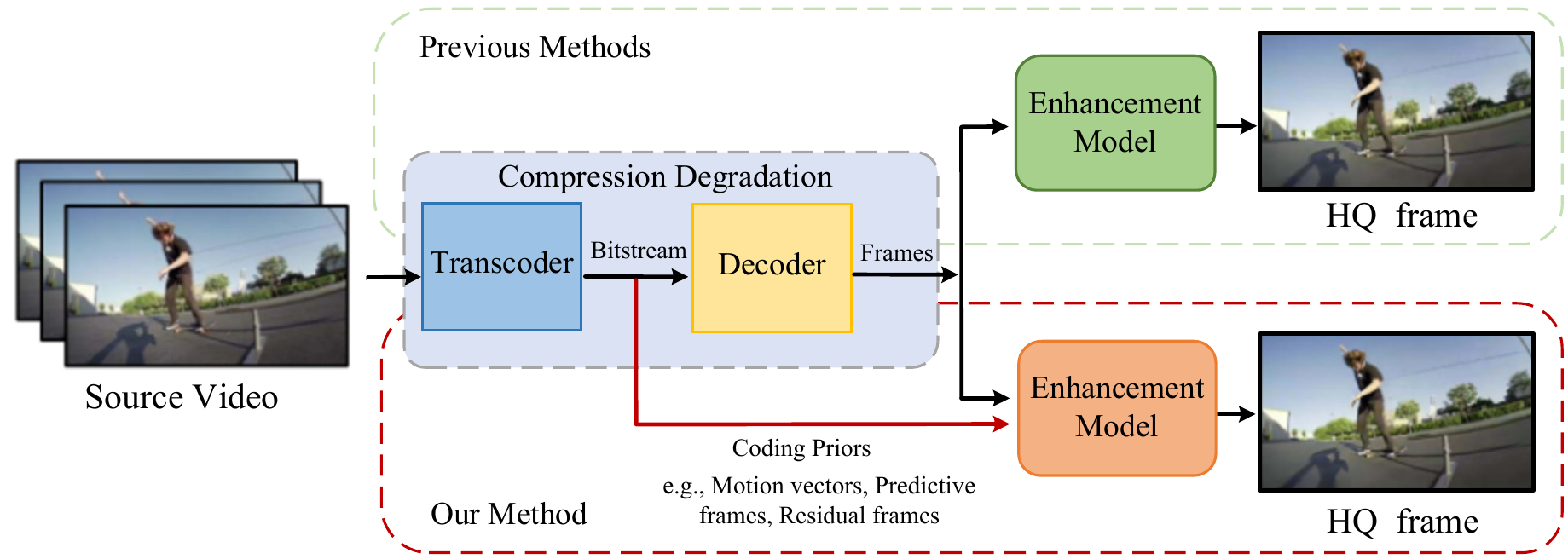}
\caption{Comparison between previous methods and our CPGA. Compared with previous methods, the coding priors are extracted from the bitstream to input into our enhancement model for VQE.}
\label{fig_data}
\vspace{-1.5em}
\end{figure}

In the past decade, extensive methods have emerged for compressed video quality enhancement (VQE), especially after the rise of deep learning. 
Inspired by image restoration~\cite{ref66,ref67,ref68} task for improving the quality of degraded images, the single-frame-based methods~\cite{ref8,ref9,ref10,ref11} are firstly proposed to achieve substantial breakthroughs by learning the inverse mapping from spatial information of low-quality (LQ) and high-quality (HQ) frames. Moreover, multi-frame-based approaches~\cite{ref19,ref22,ref23,ref24,ref25,ref26,ref27,ref65} devote to exploiting the temporal correlation between multiple consecutive LQ frames, consequently yielding superior performance on details restoration over single-frame-based methods in VQE task.
To leverage the temporal information, researchers commonly utilize optical flow~\cite{ref22,ref23} or employ deformable convolutions~\cite{ref24,ref25,ref27} to align adjacent frames toward the HQ frame to be reconstructed. 
However, the introduction of optical flow estimation inevitably increases the computational overhead, and inaccurate flow estimation and alignment may degrade the final quality~\cite{ref24}.

It is noteworthy that compressed videos inherently encompass coding information, such as motion vector (MV), predictive frame, residual frame, and partition map, collectively known as coding priors. 
MVs represent the temporal information between adjacent frames, predictive frames provide aligned results from the decoding process and the residual frame represents the difference between the current frame and the corresponding predictive frame. 
These coding priors can be extracted from the bitstream, characterizing explicit temporal and spatial information of compressed videos.

Recently, various high-level vision tasks \cite{ref56,ref57} and low-level vision tasks~\cite{ref55,ref15} have benefited from such coding priors. Meanwhile, a few approaches leveraging coding priors have yielded substantial success in compressed video super-resolution (VSR)~\cite{ref42,ref43,ref44}. Notably, compressed domain deep video super-resolution (CD-VSR) ~\cite{ref44} establishes a pioneering dataset with coding priors for the VSR task.
Despite the commendable performance achieved in these tasks, the full and effective utilization of valuable coding priors within compressed videos remains inadequately addressed in the VQE task. Additionally, existing VQE datasets typically comprise solely raw and compressed videos, lacking explicit provision of coding priors. Recognizing this, we are inspired to explore the potential of coding priors for further improving performance in VQE task.

In this paper, we firstly construct a new \textbf{V}ideo \textbf{C}oding \textbf{P}riors (VCP) dataset, comprising 300 raw videos compressed by various HEVC configurations and corresponding coding priors extracted from the bitstream, including MVs, predictive frames and residual frames. VCP dataset can remedy the shortage of previous datasets on the lack of coding priors.
%and bring improvement of performance for compressed video quality enhancement. 
Based on VCP dataset, we propose a novel coding priors-guided aggregation network, named CPGA. The CPGA consists of three modules: the inter-frame temporal aggregation (ITA) module, the multi-scale non-local aggregation (MNA) module and the quality enhancement (QE) module. Specifically, the ITA module explores the inter-frame correlations among the multiple compressed frames with the guidance of MVs and predictive frames to generate effective temporal features. 
Then, the MNA module is designed to globally capture the spatial correlations within the feature. It obtains the spatially-aggregated features with the guidance of the current residual frames.
After that, the QE module is constructed to enhance the spatially-aggregated feature to generate the HQ video frame. We illustrate the comparison between previous methods and our method in Fig. \ref{fig_data}. With the benefits of coding priors and effective feature aggregation modules, our CPGA outperforms previous methods on public testing sequences. 
The main contributions can be summarized as follows:
\begin{itemize}
\item
We establish a compressed videos with coding priors dataset for VQE task, named as VCP, which includes LQ sequences, HQ sequences and three coding priors extracted from bitstream (MVs, predictive frames and residual frames). The dataset remedies the shortage of previous datasets without coding priors in VQE task.

\item
We propose a coding priors-guided aggregation (CPGA) network for VQE task. The CPGA composed of feature aggregation modules can efficiently achieve better temporal and spatial features representation with leveraging valuable coding priors in our dataset.

\item
Experimental results demonstrate that our method achieves a performance gain of more than 0.03dB compared to previous state-of-the-art methods and outperforms \cite{ref27} by 10\% on inference speed.

\end{itemize}

\section{Related Work}
\label{sec:formatting}

\subsection{Compressed Video Quality Enhancement}
In the early years, several single-frame-based methods \cite{ref8, ref9, ref10, ref11, ref13, ref14, ref15, ref15-2} were introduced to leverage a LQ frame of compressed video to generate a HQ frame, aiming at enhancing the quality of compressed videos. 
For instance, the artifact reduction convolutional neural network \cite{ref8} employed a 4-layer CNN to reduce compression artifacts for Joint Photographic Experts Group (JPEG) images. Subsequently, 
% a denoising CNN \cite{ref15-1} was devised with a global residual learning scheme to enhance JPEG image quality. 
the residual non-local attention network \cite{ref15-2} was introduced to capture long-range dependencies between pixels for JPEG image enhancement. 
% A generative adversarial network \cite{ref15-3} was used to enhance the perceptual quality of compressed images for single-frame compressed video quality enhancement.
% Furthermore, a deep CNN-based auto-decoder \cite{ref12} designed a 10-layer CNN to reduce distortions in compressed video. Subsequently, the decoder-side scalable CNN \cite{ref12} was developed for compressed video quality enhancement, and its two subnets were utilized to reduce artifacts in intra-coding and inter-coding, respectively. 
To leverage prior information from compressed video, a partition-aware convolutional neural network \cite{ref15} was employed to utilize coding unit information from partition images to enhance the quality of compressed frames.

\begin{table*}[!t]
        \setlength{\abovecaptionskip}{0.1cm}
	\setlength{\belowcaptionskip}{0.1cm}
	\caption{Comparison between existing video enhancement datasets and our dataset.} \label{DatasetComp}
	\setlength{\tabcolsep}{2.3mm}
	\renewcommand\arraystretch{1.0}
	\fontsize{7}{9}\selectfont
	\centering
	\begin{tabular}{c|c|c|c||c||c|ccccc}
            \midrule[0.2mm]
		% \hline  
            Dataset  &  Type   & Number  & \multicolumn{2}{c|}{\makecell[c]{Resolution}}   & Compression Settings   & Coding Priors \\
		\midrule[0.2mm] % \hline  
            LDV \cite{ref33}  &  Training+Validation+Test  &  240 & \multicolumn{2}{c|}{\makecell[c]{960 $\times$ 536}}  &  { LDP  at QP=37   }  & -\\
		\midrule[0.2mm] %\hline  
            LDV2.0 \cite{ref34}  &  Training+Validation+Test  & 335 & \multicolumn{2}{c|}{\makecell[c]{4K, 960 $\times$ 536}}  &  { LDP  at QP=37 } & - \\
		\midrule[0.2mm] %\hline  
            MFQE2.0 \cite{ref23} &   Training  & 106  &   \multicolumn{2}{c|}{\makecell[c]{SIF, CIF, 720$\times$480, 4CIF, \\ 360p, 1080p and 2K }}   & { LDP  at QPs=22,27,32,37,42 } &-\\
		\midrule[0.2mm] %\hline  
            VCP (Ours)  &  Training  & 300  & \multicolumn{2}{c|}{\makecell[c]{272$\times$480, SIF, CIF, 640$\times$480, 720$\times$480, 4CIF,\\ 360p, 720p, 1080p, 2K and 4K}} & { LDB \& RA at QPs=22,27,32,37} & \checkmark  \\
		\midrule[0.2mm] % \hline
	\end{tabular}
\end{table*}

Recently, several multi-frame-based methods have been proposed for enhancing the quality of compressed videos by utilizing information from multiple frames. In these methods, optical flow is employed as motion information to align adjacent frames for quality enhancement \cite{ref22,ref23}. For instance, the multi-frame quality enhancement (MFQE) network  \cite{ref22} initially used optical flow to align adjacent compressed frames, achieving quality enhancement for compressed video. Subsequently, MFQE2.0  \cite{ref23} introduced a multi-scale approach, batch normalization and dense connection to further enhance performance.
However, in scenarios with significant motion, even a small alignment error may lead to serious artifacts in aligned frames, resulting in poor quality of the composed HR frame. 
In addition to flow-based methods, deformable convolution-based methods have been proposed \cite{ref23, ref24, ref26} to learn offsets from compressed frames to obtain aligned features for VQE. For instance, spatio-temporal deformable fusion (STDF) network \cite{ref24} jointly predicts all deformable offsets for multiple frames based on deformable convolution network (DCN)~\cite{ref28} to achieve enhancement. Furthermore, recursive fusion and deformable spatiotemporal attention (RFDA) network  \cite{ref25} employs a recursive fusion scheme to exploit relevant information from multiple frames based on DCN~\cite{ref28} in a large temporal range. The spatio-temporal detail information retrieval (STDR) network \cite{ref27} introduces a multi-path deformable alignment module to generate more accurate offsets by integrating the alignment features of different receptive fields, enabling better recovery of temporal detail information for generating HQ videos.

Although previous multi-frame-based methods achieve state-of-the-art performance, they ignore the coding information of compressed videos, limiting enhancement performance improvement. 
Such coding priors contain additional temporal and spatial information, and naturally embedded within the bitstream of compressed video. 
Recently, a few pioneering studies~\cite{ref42,ref43,ref44} have been proposed to incorporate coding priors in the VSR task. 
For instance, CD-VSR ~\cite{ref44} designed a new framework to utilize the deep priors and coding priors to achieve the improvement of performance for VSR. Besides, the codec information assisted framework \cite{ref43}  utilized the motion vector to model the temporal relationships between adjacent frames and skip the redundant pixels based on the residual frame to achieve high-efficient VSR.
Inspired by these works, we have constructed a compressed video quality enhancement dataset that includes LQ frames and their coding priors. Based on this datatset,  we design a coding priors-guided aggregation network for VQE.

% \vspace{-0.2em}
\subsection{Video Enhancement Datasets}

In the past decade, numerous datasets \cite{ref30, ref31, ref32, ref33, ref34, ref23} have been developed for compressed video quality enhancement and they consist of HQ sequences and corresponding compression configurations, such as Low Delay P (LDP) configuration and different quantization parameters (QPs). We summarize the characteristics of some representative datasets in~\cref{DatasetComp}.
Although compressed video quality enhancement methods developed based on these datasets \cite{ref23, ref33, ref34} have achieved notable success, their applicability has been largely limited in exploring video representations to further improve enhancement performance. Moreover, some coding priors (e.g., MVs, predictive frames, and residual frames) contain rich temporal and spatial information extracted from compressed videos, which can assist in constructing HQ videos. To efficiently explore the potential performance advantages offered by the coding priors of compressed videos, we propose our VCP dataset to accommodate interactions between compressed video quality enhancement and video coding.

\begin{figure}[!t]
	\centering
	\begin{minipage}[b]{0.24\linewidth}
		\centering
		\centerline{\epsfig{figure=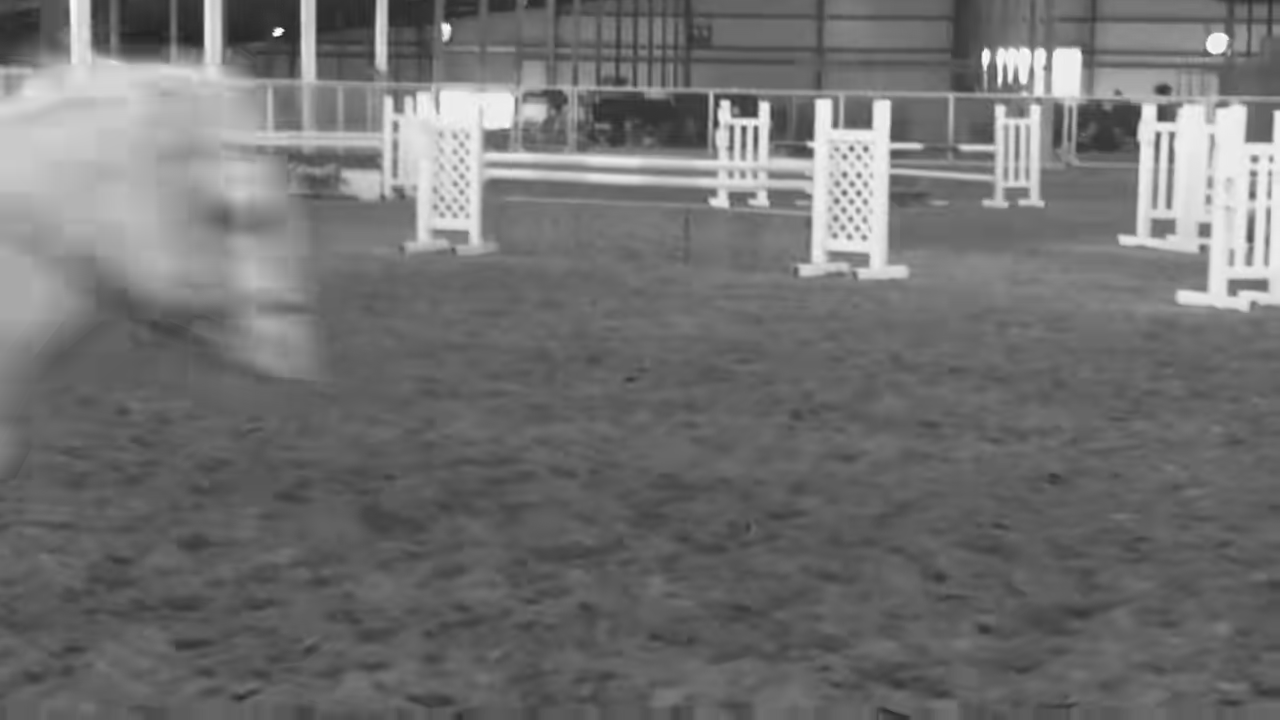,width=2.00cm}}
	\end{minipage}
	\begin{minipage}[b]{0.24\linewidth}
		\centering
		\centerline{\epsfig{figure=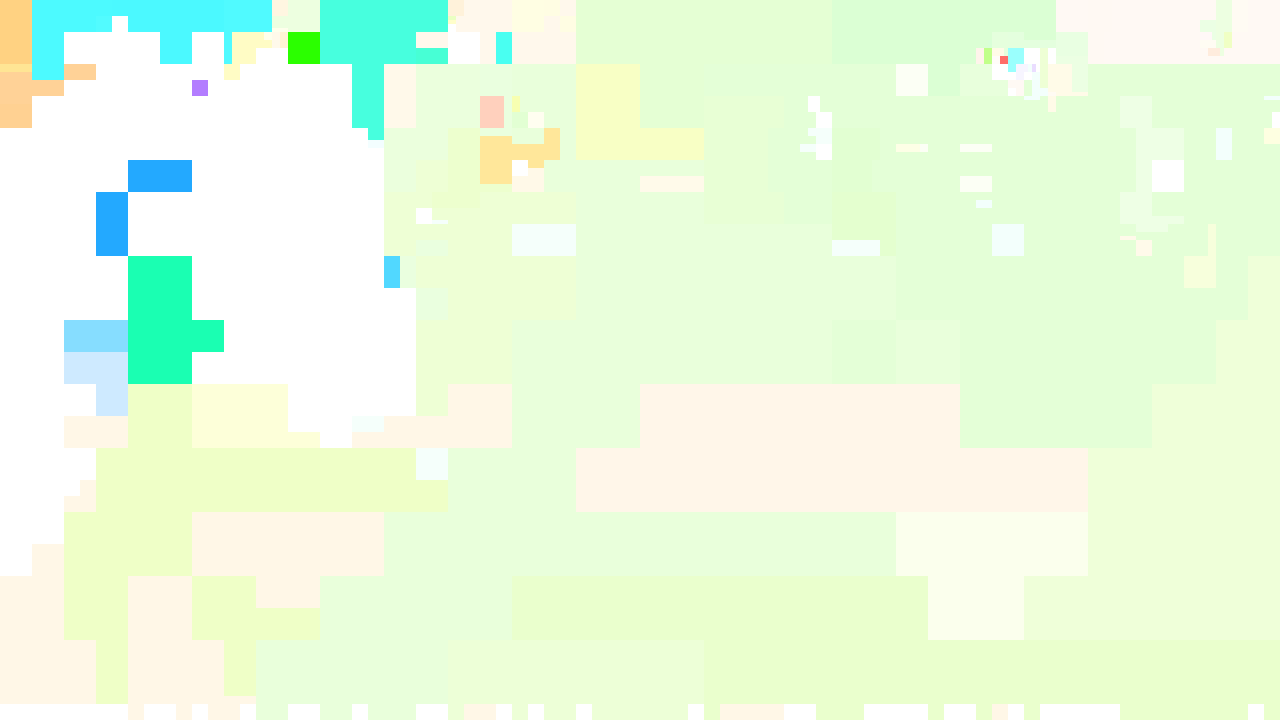,width=2.00cm}}
	\end{minipage}
	\begin{minipage}[b]{0.24\linewidth}
		\centering
		\centerline{\epsfig{figure=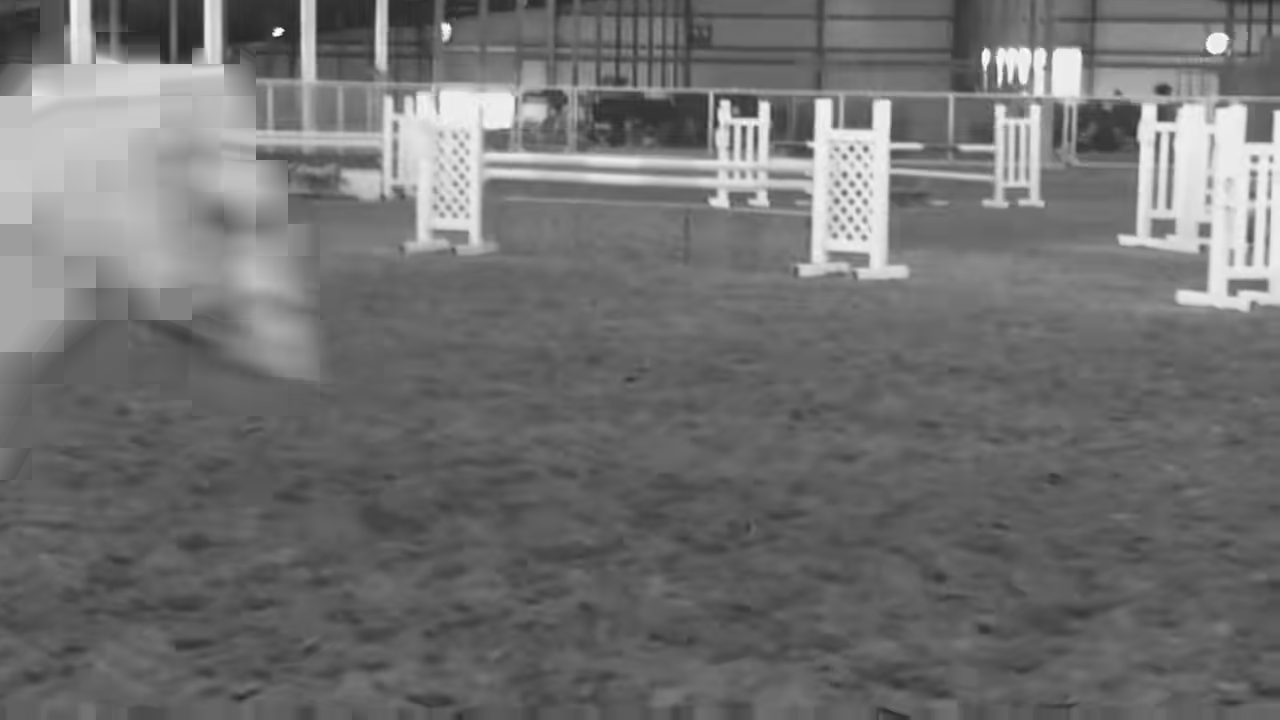,width=2.00cm}}
	\end{minipage}
        \begin{minipage}[b]{0.24\linewidth}
		\centering
		\centerline{\epsfig{figure=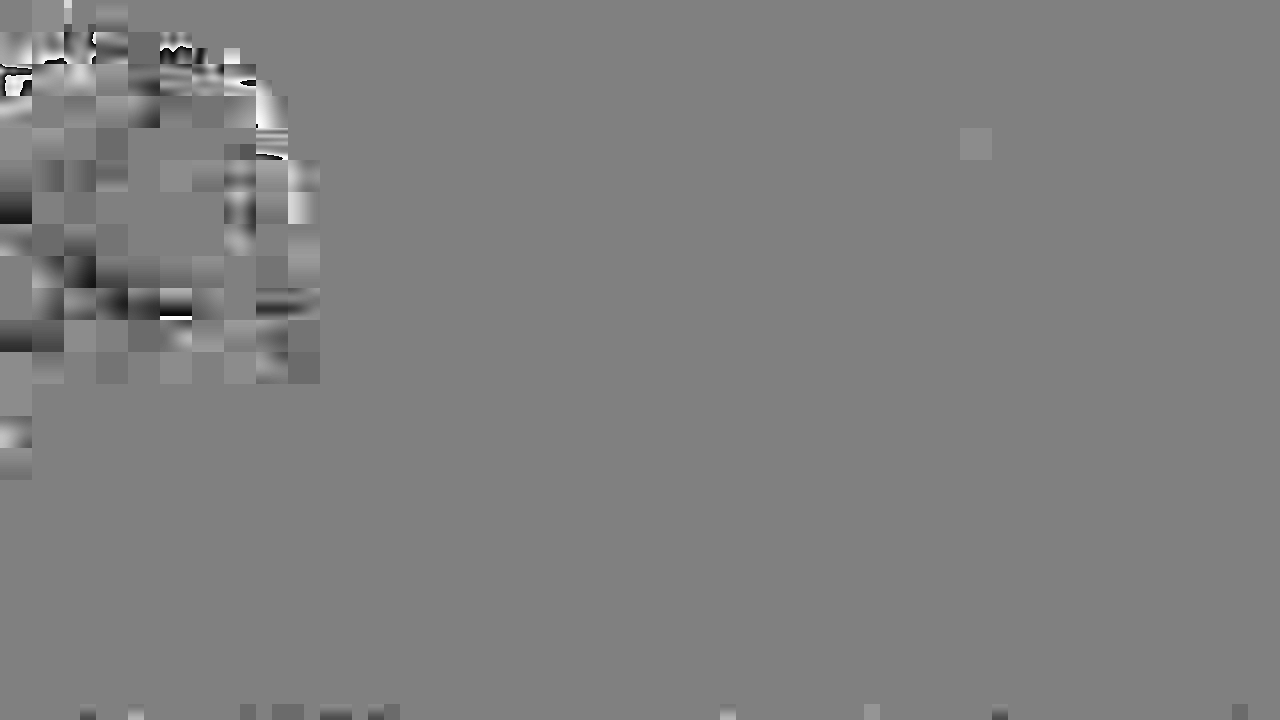,width=2.00cm}}
	\end{minipage}
	
	\begin{minipage}[b]{0.24\linewidth}
		\centering
		\centerline{\epsfig{figure=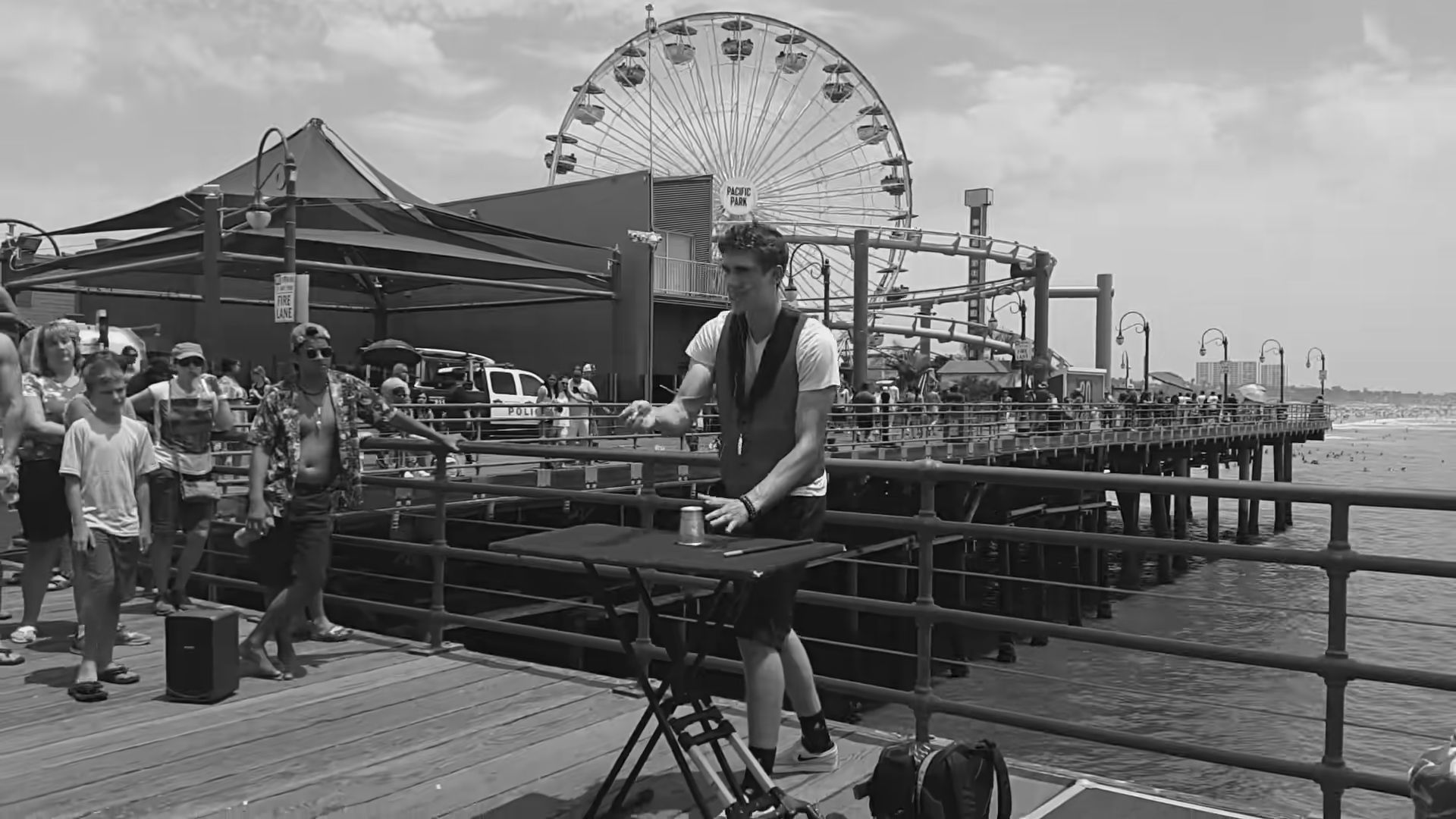,width=2.00cm}}
	\end{minipage}
	\begin{minipage}[b]{0.24\linewidth}
		\centering
		\centerline{\epsfig{figure=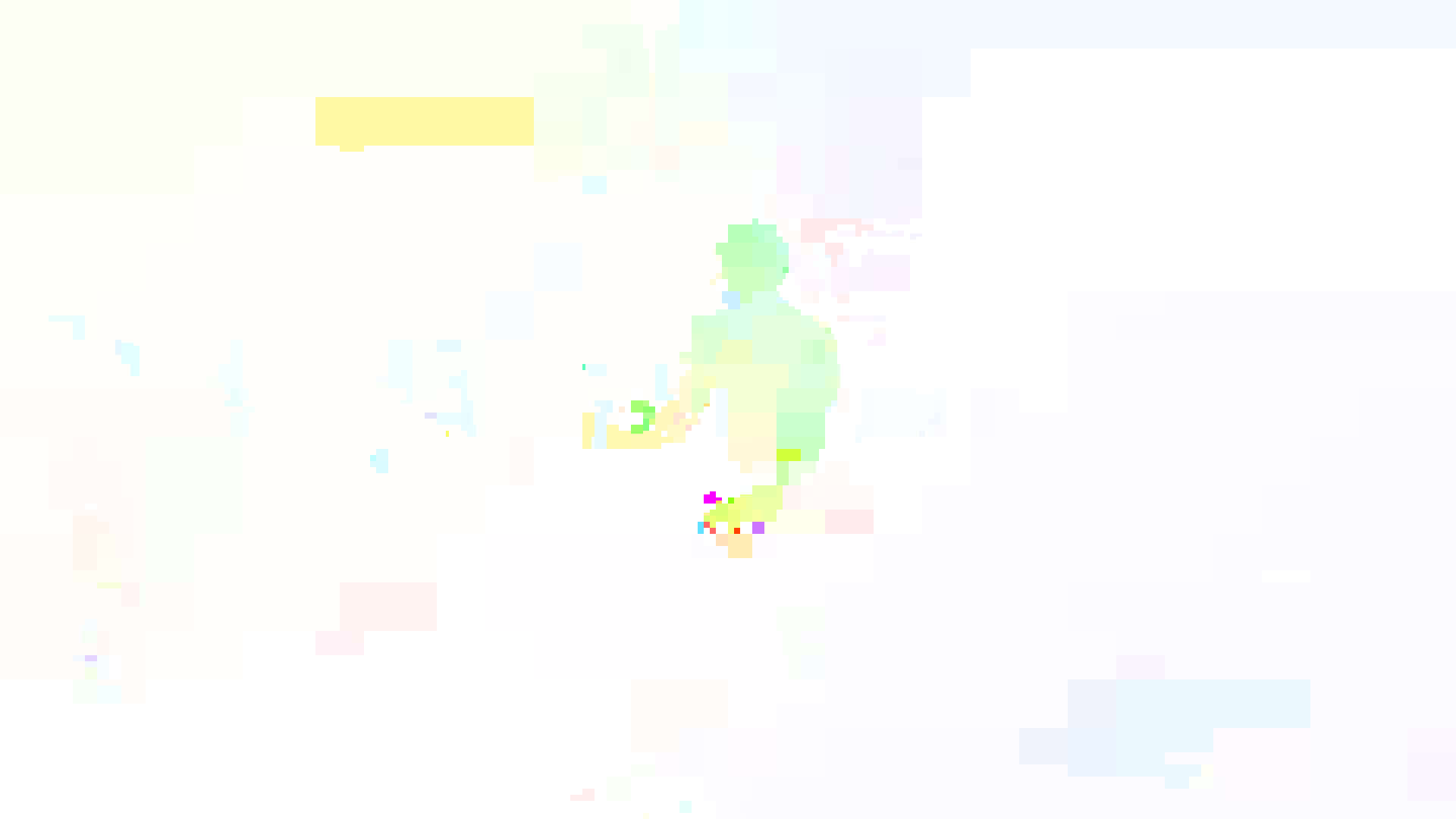,width=2.00cm}}
	\end{minipage}
	\begin{minipage}[b]{0.24\linewidth}
		\centering
		\centerline{\epsfig{figure=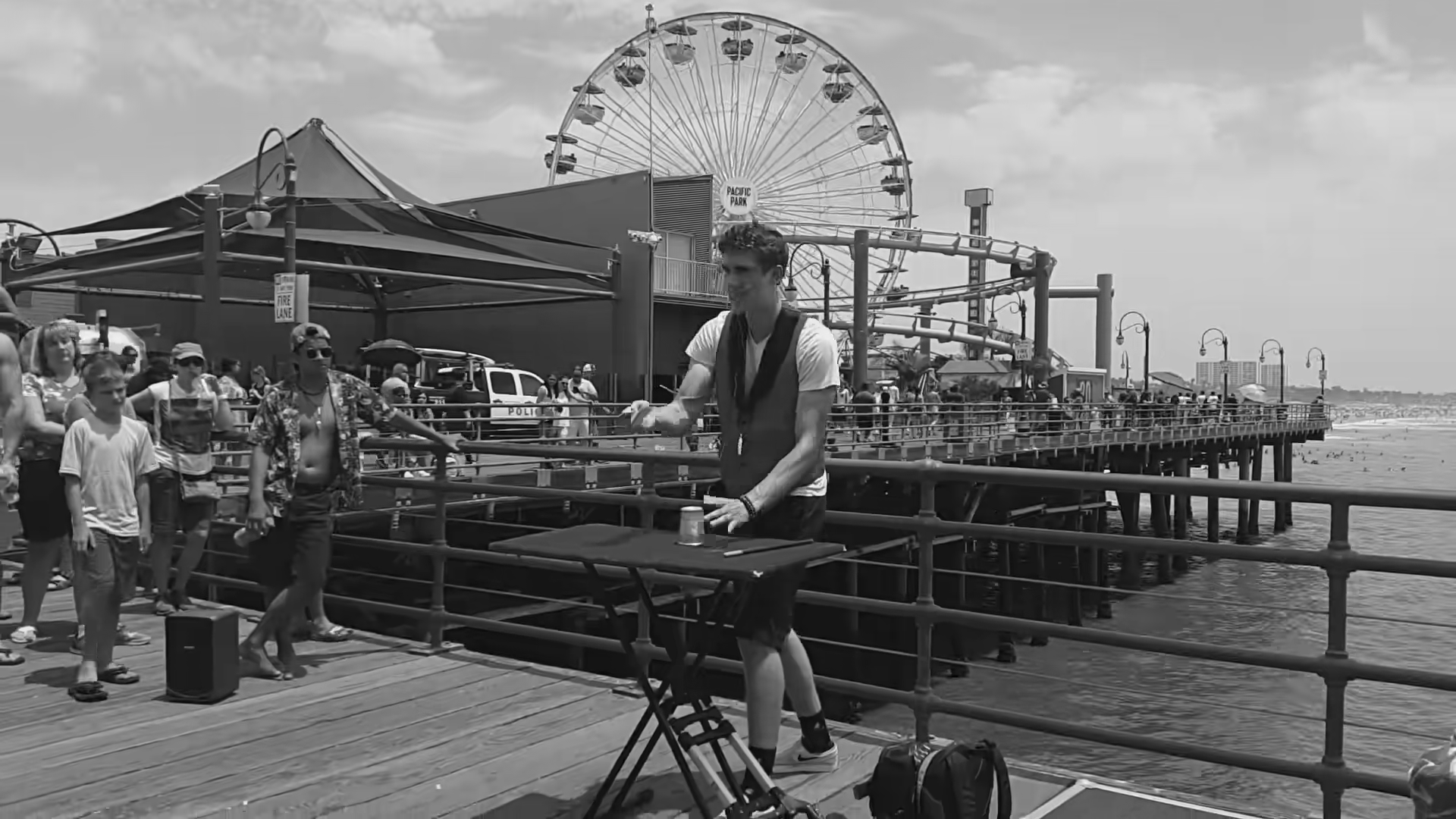,width=2.00cm}}
	\end{minipage}
        \begin{minipage}[b]{0.24\linewidth}
		\centering
		\centerline{\epsfig{figure=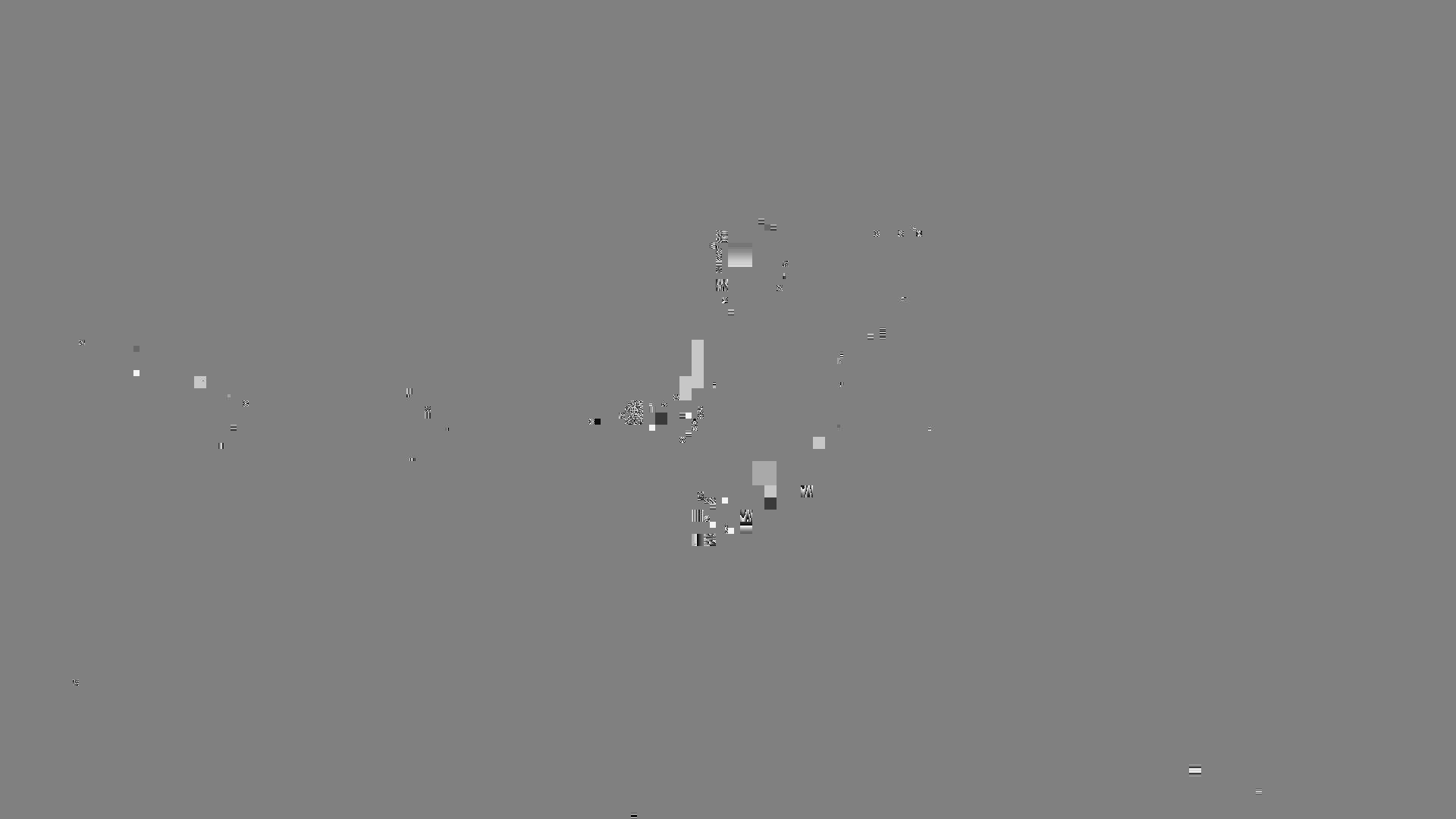,width=2.00cm}}
	\end{minipage}

 	\begin{minipage}[b]{0.24\linewidth}
		\centering
		\centerline{\epsfig{figure=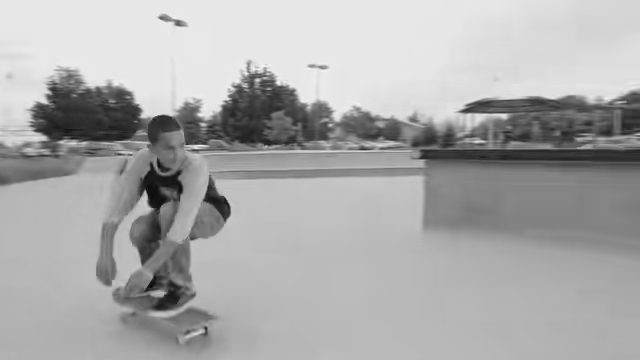,width=2.00cm}}
		\footnotesize{LQ frame  }  %\medskip
	\end{minipage}
	\begin{minipage}[b]{0.24\linewidth}
		\centering
		\centerline{\epsfig{figure=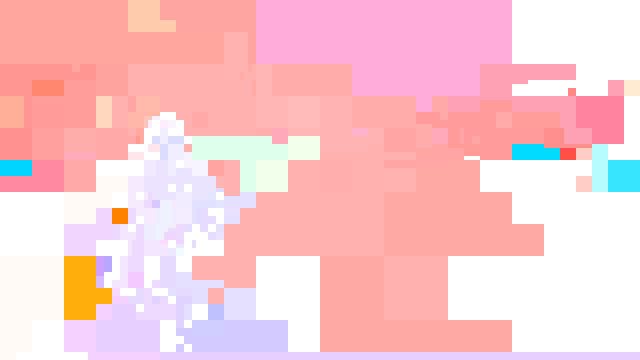,width=2.00cm}}
		\footnotesize{Motion vector }  %\medskip
	\end{minipage}
	\begin{minipage}[b]{0.24\linewidth}
		\centering
		\centerline{\epsfig{figure=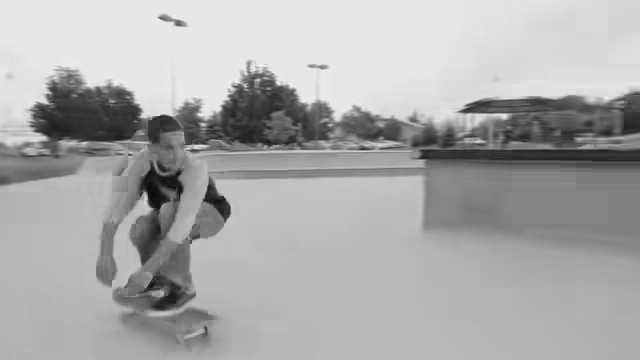,width=2.00cm}}
		\footnotesize{Predictive frame}  %\medskip
	\end{minipage}
        \begin{minipage}[b]{0.24\linewidth}
		\centering
		\centerline{\epsfig{figure=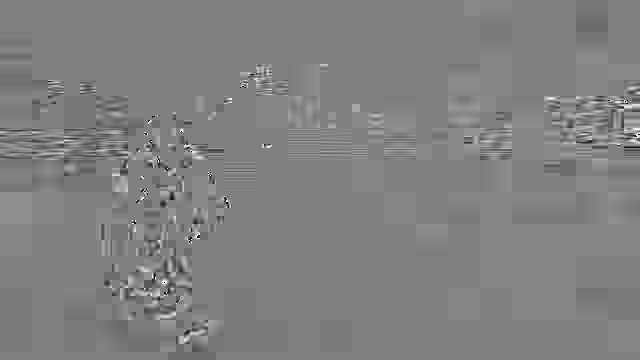,width=2.00cm}}
		\footnotesize{Residual frame  }  %\medskip
	\end{minipage}
	\caption{Some examples of  LQ frames, motion vectors, predictive frames and residual frames in proposed VCP dataset.}
	\label{exp_vision}
  \vspace{-1.0em}
\end{figure}

\begin{figure*}[!t]
\vspace{-2.8em}
	\centering
	\includegraphics[width=6.8in]{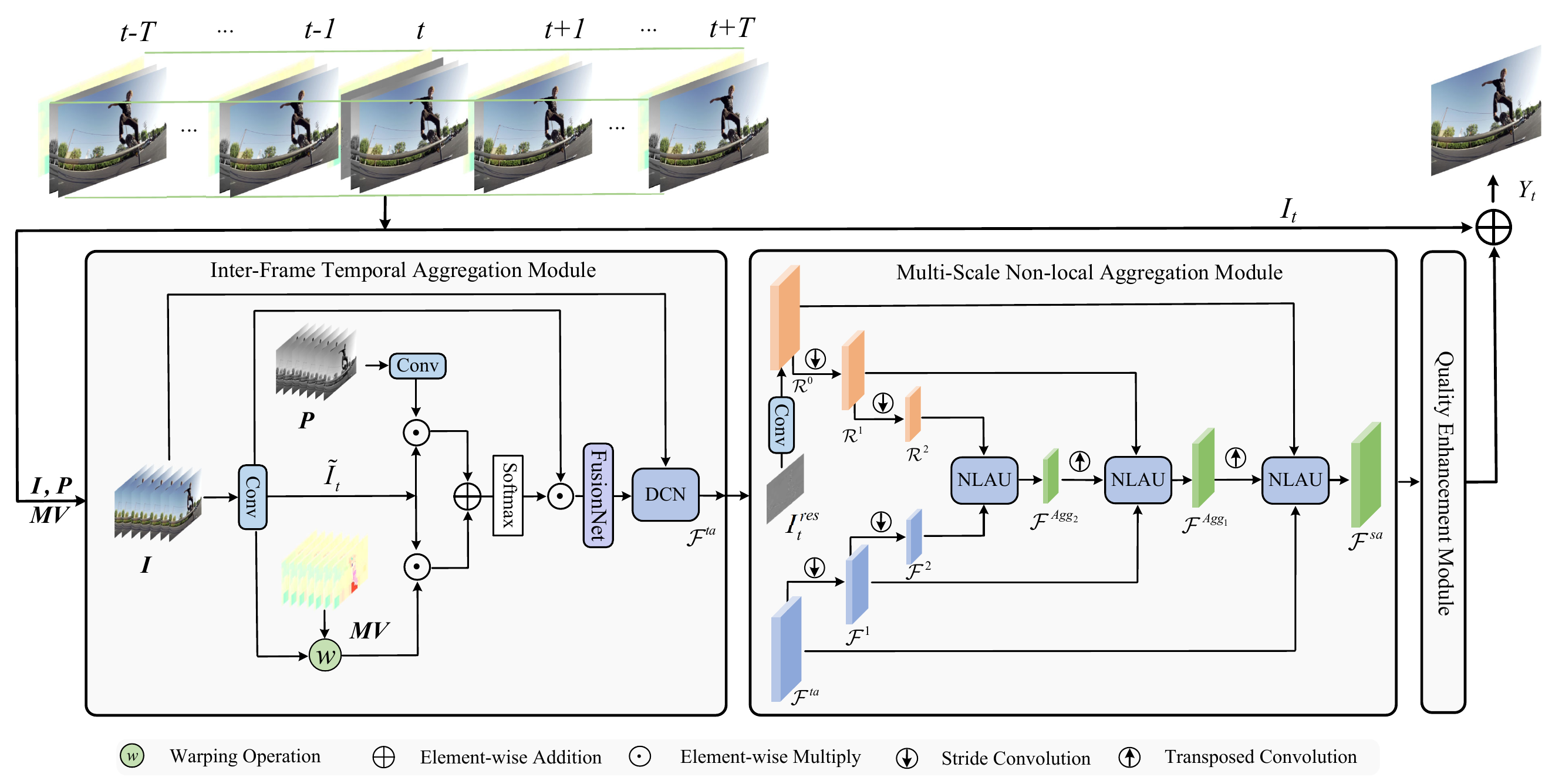}
	\caption{The architecture of CPGA. The MVs and predictive frames are fed into the ITA module to obtain the temporally-aggregated feature, and current residual frame is utilized in the MNA module to obtain the spatially-aggregated feature. }
	\label{fig_pipline}
 \vspace{-1.5em}
\end{figure*}

\vspace{-0.7em}
\section{Our dataset}
\noindent
\textbf{Data collection.}
In the compressed video quality enhancement task, existing datasets \cite{ref22, ref32, ref33} typically include only raw and compressed videos, lacking the provision of coding priors, which limits design of our network.
Herein, we collect 300 videos from CDVL \cite{ref29} and original VIMEO videos \cite{ref47}. The raw videos are selected across diverse types of content, such as wilds, urban, daily routines and professional sports. Accordingly, 300 video clips are extracted from these videos and each of them consists of 48 frames. All LQ sequences are compressed with HEVC software HM 16.25 \cite{ref1}. In particular, we adopt two configurations, i.e., Low Delay B (LDB) and Random Access (RA), and employ four QPs: 22, 27, 32 and 37, greatly facilitating the study of joint video coding and enhancement. 
Additionally, we extract three types of coding priors embedded in the bitstream of compressed videos, which include MVs, predictive frames and residual frames.  

\noindent
\textbf{Dataset analysis.}
Some examples from our VCP dataset, as shown in Fig. \ref{exp_vision}, cover a wide range of real-world scenarios. This diversity allows for a comprehensive evaluation of enhancement performance across different applications, including autonomous driving, video surveillance and photo editing. In addition to the diversity of scenarios, our VCP dataset includes various videos with different resolutions from 272 $\times$ 480 to 4K. Two coding configurations and four QPs are adopted to generate the different degradation compressed videos from raw videos, which facilitates the evaluation of the effectiveness of compressed video quality enhancement methods. 
Furthermore, the corresponding three coding priors of compressed videos also were extracted, i.e.,  MVs, predictive frames and residual frames, which provides the facilitation for the application of coding priors in VQE task. We summarize the characteristics of our dataset in~\cref{DatasetComp}. This comprehensive dataset facilitates the evaluation of the effectiveness of compressed video quality enhancement methods.

\vspace{-0.2em}
\section{Methodology}
\subsection{Overall Framework}
The architecture of our CPGA network is illustrated in~\cref{fig_pipline}. It comprises three key modules: the inter-frame temporal aggregation (ITA) module, the multi-scale non-local aggregation (MNA) module and the quality enhancement (QE) module. 
Firstly, ITA module conducts spatial feature extraction and temporal feature fusion on consecutive LQ frames and corresponding two coding priors, i.e.,  MVs and predictive frames. ITA module exploits temporal correlations of adjacent frames to generate temporally-aggregated features under the guidance of these two coding priors.
Subsequently, the temporally-aggregated features and the current residual frame are input into the MNA module, where non-local multi-scale features are integrated, resulting in the generation of spatially-aggregated features. 
Finally, the generated features are fed into the QE module to derive the ultimately enhanced feature.
The details of the proposed ITA, MNA and QE modules are explained in the following~\cref{sec:ITA,sec:MNA,sec:QE}.

\subsection{Inter-frame Temporal Aggregation Module}
\label{sec:ITA}
\vspace{-0.5em}
ITA module aims to effectively align features of adjacent frames, it incorporates two coding priors, i.e., MVs and predictive frames, to guide the fusion of features of LQ frames, thus generating the temporally-aggregated features.

Given a sequence of $2T+1$ consecutive LQ frames $\textit{\textbf{I}} = I_{[t-T:t+T]}$, where $I_t$ represents the target frame to be enhanced and the others are reference frames. Corresponding predictive frames and motion vectors are denoted as $\textit{\textbf{P}} = P_{[t-T:t+T]}$ and $\textit{\textbf{MV}} = MV_{[t-T+1:t+T]}$, note that there are $2T$ MVs in $2T+1$ frames.
We utilize a convolution layer with $3\times3$  kernel size on $I_{[t-T:t+T]}$ and $P_{[t-T:t+T]}$ to extract their features. The extracted features are denoted as $\widetilde{I}_{[t-T:t+T]}$ and $\widetilde{P}_{[t-T:t+T]}$. 
In addition, leveraging the motion vectors $MV_{[t-T+1:t+T]}$, the features of the LQ frames are aligned via a $Warp$ operation \cite{ref58} and obtain the aligned features $\widetilde{\mathcal{F}}_{[t-T+1:t+T]}$.
Since the first frame does not have the corresponding MV from the forward frame, we reuse $\widetilde{I}_{t-T}$ to represent the first aligned feature $\widetilde{\mathcal{F}}_{t-T}$.
Usually, the predictive frames are aligned results from the video decoding process while aligned features are the post-process results from motion vectors. Thus, we use these two aligned results to explore the inter-frame correlations among frames for generating temporally-aggregated features.

Specifically, the inter-frame correlations between current LQ frame feature $\widetilde{I_{t}}$, features of predictive frames $\widetilde{P}_{[t-T:t+T]}$ and aligned features $\widetilde{\mathcal{F}}_{[t-T:t+T]}$ are explored via
\begin{equation}
\hat{P}_{[t-T:t+T]}= \widetilde{I_{t}} \odot  \widetilde{P}_{[t-T:t+T]}
\end{equation}
and
\begin{equation}
\hat{\mathcal{F}}_{[t-T:t+T]} = \widetilde{I_{t}} \odot \widetilde{\mathcal{F}}_{[t-T:t+T]},
\end{equation}
where $\odot $ is an element-wise multiply operation. Then, these two temporal features are added and multiplied with the features of LQ frames for compensating the temporal information of LQ frames to obtain the temporally-compensated features, as shown in~\cref{eq:ITAtc}
\begin{equation}
\label{eq:ITAtc}
\mathcal{F}^{c}_{i} = \widetilde{I}_{i} \odot \sigma(\hat{P}_{i}+{\mathcal{\hat{F}}_{i}}),
\end{equation}
here $i\in[t-T, t+T]$ and $\sigma$ is softmax function. 
After that, we use a $FusionNet$ consisting of a convolution layer with $3\times3$ kernel size and a similar Unet\cite{ref40} from STDF~\cite{ref24}, to obtain the fused feature via fusing temporally-compensated features, as shown in~\cref{eq:ITAfusion}
\begin{equation}
\label{eq:ITAfusion}
\mathcal{F}^{f}_{t} = \text{FusionNet}(\mathcal{F}^{c}_{[t-T:t+T]}).
\end{equation}
To further aggregate temporal information, we employ a  DCN~\cite{ref28} to aggregate the temporal feature $\mathcal{F}^{f}_{t}$ to finally obtain the temporally-aggregated feature $\mathcal{F}^{ta}$.

\subsection{Multi-Scale Non-Local Aggregation Module}
\label{sec:MNA}

In video coding, residual frames refer to the difference between the frame predicted from the previous frame and the current frame. The difference is typically concentrated in essential regions such as edges and contour areas of frames. To fully aggregate the spatial information, we design the MNA module to globally explore the spatial information of features with the guidance of the current residual frame at three scales. The structure of the MNA module is shown in~\cref{fig_pipline}. 

Specifically, the current residual frame is firstly fed into a convolution layer with $3\times3$ kernel size to obtain its feature, denoted as $\mathcal{R}^{0}$. A non-local aggregation unit (NLAU) is designed to globally explore spatial information of features with the guidance of the feature $\mathcal{R}^{0}$. Then, NLAU is adopted at three different scales ($s$=$0,1,2$) to progressively aggregate features to generate spatial-aggregated feature.
We denote temporally-aggregated feature $\mathcal{F}^{ta}$ as feature $\mathcal{F}^{0}$ at scale $s$=0 in the MNA module for spatial aggregation. In detail, feature $\mathcal{F}^{0}$ and $\mathcal{R}^{0}$ are sent into the convolution layer with a stride size of $2$ to obtain corresponding downscaled features $\mathcal{F}^{1}$, $\mathcal{F}^{2}$ and $\mathcal{R}^{1}$, $\mathcal{R}^{2}$.

\begin{figure}[!t]
	\centering
	\includegraphics[width=3.2in]{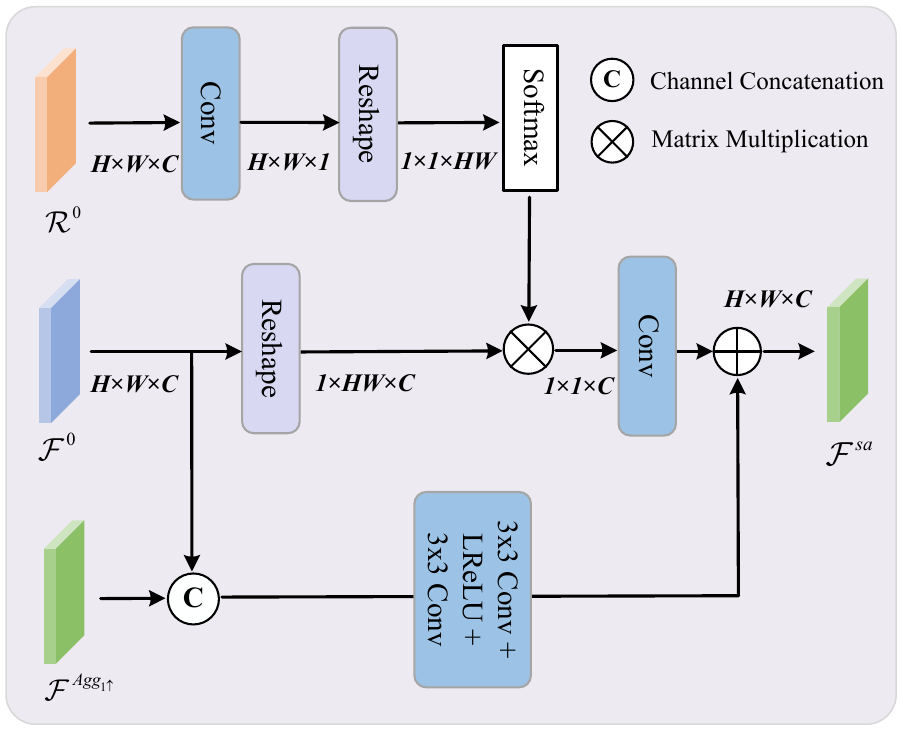}
	\caption{ The structure of non-local aggregation unit (NLAU). % Feature is aggregated with the guidance of residual feature to obtain the spatially-aggregated feature. 
 }
	\label{AU}
 \vspace{-1.0em}
\end{figure}
At scale $s$=$2$, feature $\mathcal{F}^{2}$, $\mathcal{R}^{2}$ are firstly fed into the NLAU without the previous aggregated feature to explore the spatial correlations within frames for aggregating them to obtain feature $\mathcal{F}^{Agg_{2}}$. After that, feature $\mathcal{F}^{Agg_{2}}$ is upsampled by a transposed convolution layer with a stride size of 2 to obtain the upsampled feature $\mathcal{F}^{Agg_{2\uparrow}}$.  Subsequently, the  feature $\mathcal{F}^{Agg_{2\uparrow}}$, $\mathcal{F}^{1}$, $\mathcal{R}^{1}$ are fed into the NLAU to obtain the spatially-aggregated feature $\mathcal{F}^{Agg_{1}}$ at scale $s$=1. Finally, $\mathcal{F}^{0}$, $\mathcal{R}^{0}$ and the feature $\mathcal{F}^{Agg_{1\uparrow}}$ are fed into NLAU to obtain the final spatially-aggregated feature $\mathcal{F}^{sa}$.
This process is formulated as
\begin{equation}
		\mathcal{F}^{Agg_{s}}  = \begin{cases}
		\text{NLAU}(\mathcal{F}^{s}, \mathcal{R}^{s}, \mathcal{F}^{Agg_{s+1\uparrow}}), & s = 0, 1\\
		\text{NLAU}(\mathcal{F}^{s}, \mathcal{R}^{s}), & s = 2.
	\end{cases}
\end{equation}

We illustrate the structure of NLAU at scale $s$=$0$ in~\cref{AU}. 
% Unlike traditional non-local structures~\cite{ref49,ref50,ref51}, the feature of the residual frame is employed to guide the aggregation of spatial information based on feature multiplication for generation of the spatially-aggregated feature.
In NLAU, feature $\mathcal{F}^{0}$ is transformed with the guidance of feature $\mathcal{R}^0$ to achieve the spatial aggregation within feature to obtain the spatial feature. After that, the upsampled aggregated feature $\mathcal{F}^{Agg_{1\uparrow}}$ and feature $\mathcal{F}^{0}$ 
are fused to maintain the previous aggregated information. Finally, the fused feature and the spatial feature are added to obtain the spatially-aggregated feature $\mathcal{F}^{sa}$. 

\subsection{Quality Enhancement Module}
\label{sec:QE}

\begin{figure}[!t]
        % \vspace{-2.0em}
	\centering
	\includegraphics[width=2.9in]{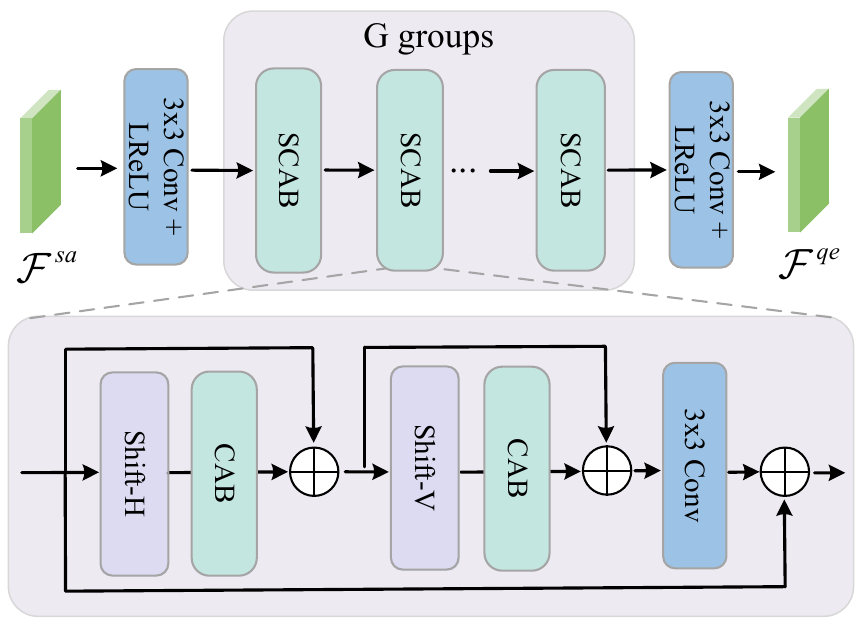}
	\caption{The structure of the quality enhancement (QE) module and the shift channel attention block (SCAB).}
	\label{QEmodule}
  \vspace{-1.5em}
\end{figure}

To further enhance spatially-aggregated feature $\mathcal{F}^{sa}$ for the composition of the high-quality frame, we first employ two convolution layers with $3\times3$ kernel size followed by the LeakyReLU activate function and then develop $G$ number of shift channel attention blocks (SCAB) to construct our QE module. The structure of the QE module and SCAB are illustrated in~\cref{QEmodule}.

In SCAB, we embed the partial channel shifting operation \cite{ref39} in the front of the channel attention block (CAB) \cite{ref48} to enlarge the receptive field.  Specifically, we choose the $\gamma$ proportion feature in the middle position of the feature to use the partial channel shifting operations to enlarge the receptive field for feature enhancement without increasing parameters and complexity. We design two partial channel shifting operations with different directions, i.e., partial channel shifting operation along the horizontal direction ($Shift$-$H$) with $\hat{h}$ pixel and partial channel shifting operation along the vertical direction ($Shift$-$V$) with $\hat{w}$ pixel. The $Shift$-$H$ operation is first adopted to enlarge the receptive field along the horizontal direction at the front of CAB. Subsequently, the $Shift$-$V$ operation is adopted to enlarge the receptive field along the vertical direction at the front of the CAB. Based on these two shift operations and CABs, the SCAB is designed to construct our QE module for feature enhancement.  

\section{ Experiments}
\subsection{Experimental Settings}

We adopt our VCP dataset as our training dataset. Following \cite{ref24,ref25,ref26,ref27}, 18 standard testing sequences from common test conditions of JCT-VC \cite{ref36} are used as our testing sequences. All testing sequences are processed under LDB and RA configurations with four QPs: 22, 27, 32, and 37, to generate the compressed videos and corresponding coding priors.
The peak signal-to-noise ratio (PSNR) and structural similarity index (SSIM) \cite{ref34} are adopted for performance evaluation. The enhanced results are evaluated in terms of $\Delta$PSNR and $\Delta$SSIM, which measures the PSNR/SSIM gap between the enhanced and the compressed videos. These two metrics are offered over the Y channel of YCbCr space as previous works \cite{ref24,ref25,ref26,ref27} did.

\begin{table*}[!t]
    \vspace{-2.5em}
	\caption{Quantitative comparison in terms of $\Delta$PSNR (dB) and $\Delta$SSIM ($\times$$10^{-2}$) under LDB configuration.
 \R{Red} text indicates the best and \B{blue} text indicates the second best performance. * Video resolutions: Class A (2560$\times$1600), Class B (1920$\times$1080), Class C (832$\times$480), Class D (416$\times$240) and Class E (1280$\times$720). } \label{RST_LDB}
	\setlength{\tabcolsep}{2.1mm}
	\renewcommand\arraystretch{1.0}
	\fontsize{7}{9}\selectfont
	\centering
	% \begin{threeparttable}
		\begin{tabular}{rcccc|cc|cc|cc|cc|cccc}
			% \toprule[0.2mm]
			\toprule[0.2mm]
                % \hline
			\multirow{2}{*}{\textbf{QP}}     &\multirow{2}{*}{\textbf{Class}*}   & \multirow{2}{*}{\textbf{Sequence}}     & \multicolumn{2}{c}{\textbf { MFQE }}  & \multicolumn{2}{c}{\textbf { STDF-R3L }} & \multicolumn{2}{c}{\textbf { RFDA }} & \multicolumn{2}{c}{\textbf { CF-STIF }} & \multicolumn{2}{c}{\textbf { STDR }}  &\multicolumn{2}{c}{\textbf { Ours }} \\
		\cmidrule(r){4-5}   \cmidrule(r){6-7}    \cmidrule(r){8-9}  \cmidrule(r){10-11}  \cmidrule(r){12-13}  \cmidrule(r){14-15}  \cmidrule(r){16-17}  % \cmidrule(r){18-19}
			&  &  & PSNR & SSIM  & PSNR & SSIM  & PSNR & SSIM    & PSNR & SSIM  & PSNR & SSIM  & PSNR & SSIM  \\
			\hline
			\multirow{19}{*}{37}
			& \multirow{2}{*}{A}
			& { Traffic }  & 0.36  & 0.75   & 0.63 & 1.11 & 0.68 & 1.18 & 0.67 & 1.19 & 0.71 & 1.23 & 0.73 & 1.21  \\
			\cline{3-17}  & & { PeopleOnStreet }  & 0.56  & 1.13  & 1.05 & 1.89 & 1.15 & 1.98 & 1.12 & 2.00 & 1.17 & 2.04 & 1.27 & 2.12  \\
			\cline{2-17}  & & { Kimono } & 0.21  & 0.55   & 0.64 & 1.22 & 0.68 & 1.37 & 0.68 & 1.28 & 0.72 & 1.32 & 0.79 & 1.43  \\
			\cline{3-17}
			& \multirow{3}{*}{B}   & { ParkScene } & 0.18  & 0.57   & 0.46 & 1.18 & 0.53 & 1.30 & 0.49 & 1.33 & 0.56 & 1.37 & 0.56 & 1.43  \\
			\cline{3-17} &  & { Cactus }  & 0.26  & 0.63   & 0.66 & 1.29 & 0.69 & 1.27 & 0.71 & 1.39 & 0.73 & 1.38 & 0.74 & 1.39  \\
			\cline{3-17} & & { BQTerrace }  & 0.20  & 0.45   & 0.55 & 0.97 & 0.55 & 0.96 & 0.56 & 1.00 & 0.59 & 1.08 & 0.59 & 1.08 \\
			\cline{3-17} & & { BasketballDrive }  & 0.27  & 0.56   & 0.65 & 1.09 & 0.70 & 1.15 & 0.70 & 1.21 & 0.74 & 1.28 & 0.78 & 1.26  \\
			\cline{2-17} & & { RaceHorses } & 0.22  & 0.57   & 0.49 & 1.29 & 0.51 & 1.26 & 0.56 & 1.32 & 0.58 & 1.36 & 0.62 & 1.57 \\
			\cline{3-17}
			& \multirow{2}{*}{C}  & { BQMall } & 0.21  & 0.77   & 0.84 & 1.76 & 0.90 & 1.79 & 0.90 & 1.88 & 0.97 & 1.96 & 0.97 & 1.91  \\
			\cline{3-17}  & & { PartyScene }  & -0.12  & 0.34   & 0.53 & 1.76 & 0.58 & 1.77 & 0.62 & 1.84 & 0.63 & 1.91 & 0.61 & 2.02  \\
			\cline{3-17}  & & { BasketballDrill } & 0.24  & 0.75   & 0.69 & 1.37 & 0.73 & 1.37 & 0.77 & 1.51 & 0.78 & 1.57 & 0.78 & 1.46  \\
			\cline{2-17}  & & { RaceHorses }  & 0.31  & 0.86   & 0.70 & 1.87 & 0.75 & 1.92 & 0.78 & 2.01 & 0.83 & 2.10  & 0.85 & 2.23  \\
			\cline{3-17}
			& \multirow{2}{*}{D}   & { BQSquare } & -0.53  & -0.29   & 0.69 & 1.14  & 0.82 & 1.36 & 0.83 & 1.41 & 0.87 & 1.41 & 0.89 & 1.43 \\
			\cline{3-17}  & & { BlowingBubbles }  & 0.12  & 0.77   & 0.57 & 2.07 & 0.68 & 2.28 & 0.66 & 2.31 & 0.68 & 2.41 & 0.67 & 2.43 \\
			\cline{3-17}  & & { BasketballPass }   & 0.27  & 0.79  & 0.82 & 1.76 & 0.86 & 1.75 & 0.92 & 1.99 & {0.97} & {2.08} & {0.98} & {2.10} \\
			\cline{2-17}  & & { FourPeople }  &  0.46  & 0.77  & 0.90 & 1.17 & 0.96 & 1.26 & 0.96 & 1.23 & {1.02} & {1.26} & {1.04} & {1.28} \\
			\cline{3-17}
			& \multirow{2}{*}{E}  & { Johnny }   & 0.39  & 0.37  & 0.72 & 0.69 & 0.70 & 0.55 & {0.81} & {0.81} & {0.81} & {0.78} & {0.80} & {0.84} \\
			\cline{3-17}  & & { KristenAndSara }   & 0.40  & 0.55  & 0.87 & 0.86 & 0.94 & 0.91 & 0.92 & 0.92 & {0.95} & {0.96} & {1.01} & {0.93} \\
			\cline{2-17} & & \textbf{ Average }   & 0.20  & 0.56   & {0.69} & {1.36} & {0.74} & {1.41} & {0.76} & {1.48} & \textbf{\blue{0.79}} & \textbf{\blue{1.52}} & \textbf{\red{0.82}} & \textbf{\red{1.56}}  \\
			\cline{1-17}
			\multirow{1
   }{*}{{32}} &  & \textbf{ Average }  & 0.15  & 0.34   & {0.71}& {0.99} & {0.76} & {1.03} & {0.78} & {1.05} & \textbf{\blue{0.80}} & \textbf{\blue{1.07}} & \textbf{\red{0.83}} & \textbf{\red{1.10}}  \\
			\cline{1-17}
			\multirow{1}{*}{{27}}& & \textbf{ Average }  & 0.16  & 0.23  & {0.67} & {0.63} & {0.69} & {0.64} & {0.72} & {0.67} & \textbf{\blue{0.74}} & \textbf{\blue{0.70}} & \textbf{\red{0.77}} & \textbf{\red{0.75}}  \\
			\cline{1-17}
			\multirow{1}{*}{{22}}& & \textbf{ Average }  & 0.20  & 0.13   & {0.57}& {0.32} & {0.56} & {0.34} & {0.58} & {0.35} & \textbf{\blue{0.60}} & \textbf{\blue{0.37}} & \textbf{\red{0.62}} & \textbf{\red{0.39}} \\
			\toprule[0.2mm]
			% \toprule[0.2mm]
		\end{tabular}
		% \begin{tablenotes}
			% \footnotesize
			% \item[*] Video resolutions: Class A (2560$\times$1600), Class B (1920$\times$1080), Class C (832$\times$480), Class D (416$\times$240) and Class E (1280$\times$720).
		% \end{tablenotes}
	% \end{threeparttable}
 \vspace{-1.5em}
\end{table*}

\begin{table}[!t]
        \setlength{\abovecaptionskip}{0.1cm}
	\setlength{\belowcaptionskip}{0.1cm}
	\caption{Quantitative comparison in terms of $\Delta$PSNR (dB) and $\Delta$SSIM ($\times$$10^{-2}$)  under RA configuration. \R{Red} text indicates the best and \B{blue} text indicates the second best performance.} \label{RST_RA}
    \setlength{\tabcolsep}{0.3mm}
	\renewcommand\arraystretch{1.2}
	\fontsize{7}{9}\selectfont
	\centering
	\begin{tabular}{c|c|c|c|c|c|cccc}
		\cline{1-7}
		\multirow{2}{*}{\textbf{QP}}   & \multicolumn{1}{c|}{\textbf { MFQE }}  & \multicolumn{1}{c|}{\textbf { STDF-R3L }} & \multicolumn{1}{c|}{\textbf { RFDA }} & \multicolumn{1}{c|}{\textbf { CF-STIF }} & \multicolumn{1}{c|}{\textbf { STDR }}  &\multicolumn{1}{c}{\textbf { Ours }} \\
		\cline{2-7}
		&  PSNR/SSIM  & PSNR/SSIM  & PSNR/SSIM  & PSNR/SSIM  & PSNR/SSIM  & PSNR/SSIM   \\
		\hline
		{37}  & 0.11/0.41   & {0.43}/{0.78} & {0.48}/{0.90} & {0.53}/{1.02} & \textbf{\blue{0.55}}/\textbf{\blue{1.04}} & \textbf{\red{0.57}}/\textbf{\red{1.07}}  \\
		\cline{1-7}
		{32}   & 0.12/0.29   & {0.44}/{0.55} & {0.46}/{0.58}& {0.49}/{0.60} & \textbf{\blue{0.50}}/\textbf{\blue{0.63}} & \textbf{\red{0.55}}/\textbf{\red{0.67}}   \\
		\cline{1-7}
		{27} & 0.10/0.20  & {0.41}/{0.31} & {0.42}/{0.33} & {0.45}/{0.40} & \textbf{\blue{0.48}}/\textbf{\blue{0.41}} & \textbf{\red{0.53}}/\textbf{\red{0.45}}  \\
		\cline{1-7}
		{22} & 0.14/0.12   & {0.33}/{0.20} & {0.36}/{0.23} & {0.42}/{0.24} & \textbf{\blue{0.46}}/\textbf{\blue{0.26}} & \textbf{\red{0.48}}/\textbf{\red{0.27}}  \\
		\toprule[0.2mm]
	\end{tabular}
        \vspace{-1.5em}
\end{table}

\subsection{Implementation Details}

% The proposed models are developed based on Pytorch and trained on two NVIDIA GeForce RTX 3090 GPUs and Intel(R) Xeon(R) Gold 6133 CPU. 
In the training phase, we randomly crop 128 $\times$128 clips from raw videos, compressed videos and their corresponding coding priors as training samples with setting the batch size to 32. Each video clip contains 7 frames, i.e., $T$=3. We also employ data augmentation strategies (i.e., selection or flipping) to expand our dataset. We train all the models using Adam \cite{ref28} optimizer with $\beta_{1}$=0.9, $\beta_{2}$=0.999, the compensation factor $\varepsilon=1\times10^{-8}$ and the learning rate is initially set to 1$\times$${10}^{-4}$. We adopt the Charbonnier loss \cite{ref11} to train our proposed model.
For setting of our model, we set $G$=2 in the  QE module, $\hat{h}$ = $\hat{w}$ =  2 and  $\gamma$  is set to 1/8 in SCAB.  The kernel size of deformable convolution in the ITA module is set to $3\times3$. 
% Two standard objective quality metrics, i.e., $\Delta$PSNR  and $\Delta$SSIM, are adopted to evaluate the enhancement performance. 

\subsection{Comparison with State-of-the-art Methods}

To demonstrate the effectiveness of our method, we compare our method with five multi-frame-based compressed video quality enhancement methods, i.e., MFQE \cite{ref22}, STDF-R3L \cite{ref24}, RFDA \cite{ref25}, coarse-to-fine
spatio-temporal information fusion (CF-STIF) \cite{ref26} and STDR \cite{ref27}.  We retrain these methods on our dataset using their settings, the results of these methods under LDB and RA configurations are provided in~\cref{RST_LDB} and~\cref{RST_RA}, respectively. 

\noindent
\textbf{Quantitative Results.}
We present the quantitative results in terms of average $\Delta$PSNR and $\Delta$SSIM in~\cref{RST_LDB} and~\cref{RST_RA}. It can be observed from~\cref{RST_LDB} and~\cref{RST_RA} that our method consistently outperforms all the compared methods, which demonstrates that our method achieves the state-of-the-art performance under LDB and RA configurations with all different QPs. 
Specifically, our method outperforms STDF-R3L and RFDA by 0.13dB and 0.08dB in terms of average $\Delta$PSNR over 18 standard testing sequences under LDB configuration at QP = 37. Besides, compared with a state-of-the-art method, i.e., STDR, our method also achieves an improvement of 0.03dB under LDB configuration at QP = 37. Moreover, our model improves about 0.02dB-0.05dB in terms of average $\Delta$PSNR under RA configuration.

\begin{figure}[!t]
        \setlength{\abovecaptionskip}{0.1cm}
	\setlength{\belowcaptionskip}{0.1cm}
        \begin{minipage}[b]{1\linewidth}
		\centering
		\centerline{\epsfig{figure=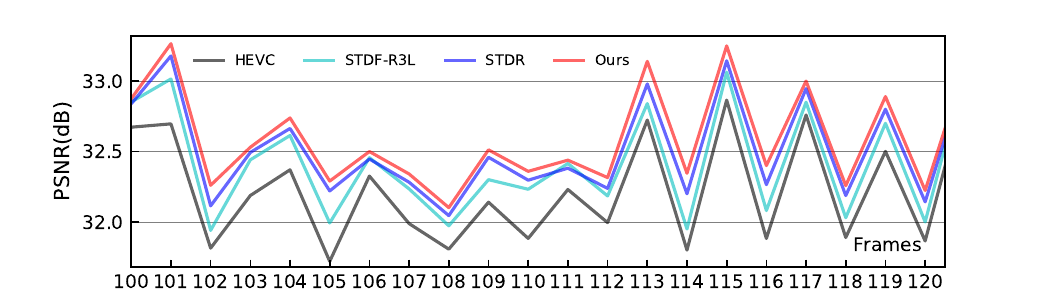,width=8.5cm}}
	\end{minipage}
	\caption{PSNR curves of HEVC, STDF-R3L, STDR and our model on BasketballPass testing sequence under LDB configuration at QP=37.}
	\label{flowf_vision}
 \vspace{-1.2em}
\end{figure}

\noindent
\textbf{Quality Fluctuation.}
Quality fluctuation is another observable measurement for the overall quality of enhanced videos. Drastic quality fluctuation of frames accounts for severe texture shaking and degradation of the quality of experience. We provide the PSNR curves of the HEVC, STDF-R3L, STDR and our method on a testing sequence, i.e., BasketballPass, in~\cref{flowf_vision}. It can be seen in~\cref{flowf_vision} that our method has not only more improvement in performance but also smaller fluctuations. 
% Therefore, our method is beneficial to mitigating the quality fluctuation of enhanced video while enhancing video quality.

\begin{figure*}[!t]
        \vspace{-1.5em}
	\centering
	\begin{minipage}[b]{0.195\linewidth}
		\centering
		\centerline{\epsfig{figure=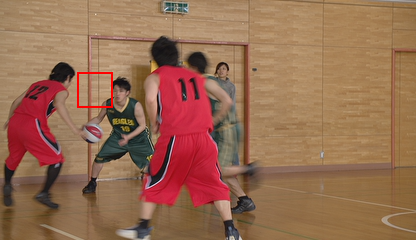,width=3.28cm}}
		\footnotesize{BasketballPass\_095 \ }  %\medskip
	\end{minipage}
	\begin{minipage}[b]{0.11\linewidth}
		\centering
		\centerline{\epsfig{figure=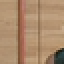,width=1.9cm}}
		\footnotesize{ \ \\ } %\medskip
	\end{minipage}
	\begin{minipage}[b]{0.11\linewidth}
		\centering
		\centerline{\epsfig{figure=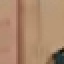,width=1.9cm}}
		\footnotesize{  \ \\  }%\medskip
	\end{minipage}
	\begin{minipage}[b]{0.11\linewidth}
		\centering
		\centerline{\epsfig{figure=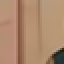,width=1.9cm}}
		\footnotesize{ \ \\  } %\medskip
	\end{minipage}
	\begin{minipage}[b]{0.11\linewidth}
		\centering
		\centerline{\epsfig{figure=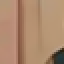,width=1.9cm}}
		\footnotesize{ \ \\  }%\medskip
	\end{minipage}
	\begin{minipage}[b]{0.11\linewidth}
		\centering
            \centerline{\epsfig{figure=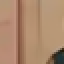,width=1.9cm}}
		\footnotesize{ \ \\  }%\medskip
	\end{minipage}
	\begin{minipage}[b]{0.11\linewidth}
		\centering
		\centerline{\epsfig{figure=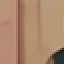,width=1.9cm}}
		\footnotesize{ \  \\  }%\medskip
	\end{minipage}
	\begin{minipage}[b]{0.11\linewidth}
		\centering
		\centerline{\epsfig{figure=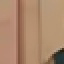,width=1.9cm}}
		\footnotesize{ \ \\  }%\medskip
	\end{minipage}
	
	\begin{minipage}[b]{0.195\linewidth}
		\centering
		\centerline{\epsfig{figure=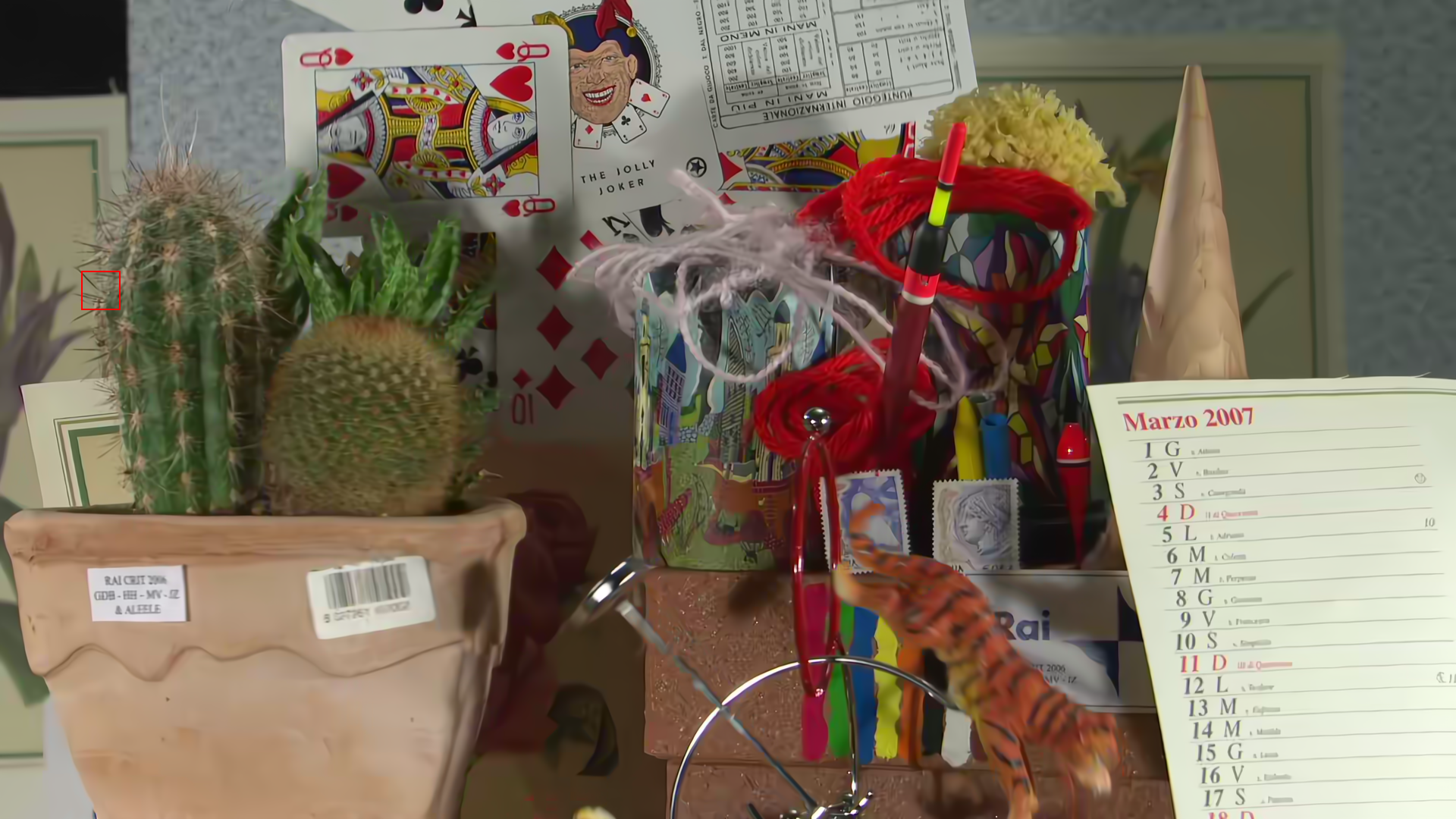,width=3.32cm}}
		\footnotesize{Cactus\_016\ }  %\medskip
	\end{minipage}
	\begin{minipage}[b]{0.11\linewidth}
		\centering
		\centerline{\epsfig{figure=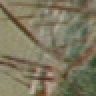,width=1.9cm}}
		\footnotesize{ \ \\  } %\medskip
	\end{minipage}
	\begin{minipage}[b]{0.11\linewidth}
		\centering
		\centerline{\epsfig{figure=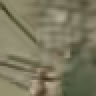,width=1.9cm}}
		\footnotesize{ \  \\  }%\medskip
	\end{minipage}
	\begin{minipage}[b]{0.11\linewidth}
		\centering
		\centerline{\epsfig{figure=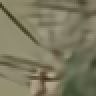,width=1.9cm}}
		\footnotesize{ \ \\  }%\medskip
	\end{minipage}
	\begin{minipage}[b]{0.11\linewidth}
		\centering
		\centerline{\epsfig{figure=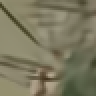,width=1.9cm}}
		\footnotesize{  \ \\  }%\medskip
	\end{minipage}
	\begin{minipage}[b]{0.11\linewidth}
		\centering
		\centerline{\epsfig{figure=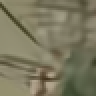,width=1.9cm}}
		\footnotesize{  \ \\  }%\medskip
	\end{minipage}
	\begin{minipage}[b]{0.11\linewidth}
		\centering
		\centerline{\epsfig{figure=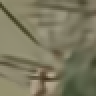,width=1.9cm}}
		\footnotesize{  \ \\  }%\medskip
	\end{minipage}
	\begin{minipage}[b]{0.11\linewidth}
		\centering
		\centerline{\epsfig{figure=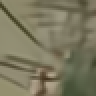,width=1.9cm}}
		\footnotesize{  \ \\  }%\medskip
	\end{minipage}
	
	\begin{minipage}[b]{0.195\linewidth}
		\centering
		\centerline{\epsfig{figure=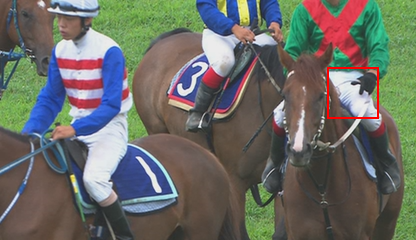,width=3.28cm}}
		\footnotesize{RaceHorses\_019    }  %\medskip
	\end{minipage}
	\begin{minipage}[b]{0.11\linewidth}
		\centering
		\centerline{\epsfig{figure=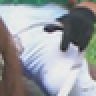,width=1.9cm}}
		\footnotesize{\  \\  } %\medskip
	\end{minipage}
	\begin{minipage}[b]{0.11\linewidth}
		\centering
		\centerline{\epsfig{figure=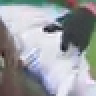,width=1.9cm}}
		\footnotesize{ \  \\  }%\medskip
	\end{minipage}
	\begin{minipage}[b]{0.11\linewidth}
		\centering
		\centerline{\epsfig{figure=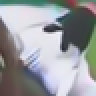,width=1.9cm}}
		\footnotesize{ \ \\  }%\medskip
	\end{minipage}
	\begin{minipage}[b]{0.11\linewidth}
		\centering
		\centerline{\epsfig{figure=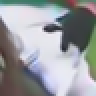,width=1.9cm}}
		\footnotesize{  \ \\  }%\medskip
	\end{minipage}
	\begin{minipage}[b]{0.11\linewidth}
		\centering
		\centerline{\epsfig{figure=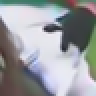,width=1.9cm}}
		\footnotesize{  \ \\  }%\medskip
	\end{minipage}
	\begin{minipage}[b]{0.11\linewidth}
		\centering
		\centerline{\epsfig{figure=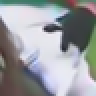,width=1.9cm}}
		\footnotesize{  \ \\  }%\medskip
	\end{minipage}
	\begin{minipage}[b]{0.11\linewidth}
		\centering
		\centerline{\epsfig{figure=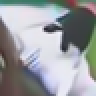,width=1.9cm}}
		\footnotesize{  \ \\  }%\medskip
	\end{minipage}
	
	\begin{minipage}[b]{0.195\linewidth}
		\centering
		\centerline{\epsfig{figure=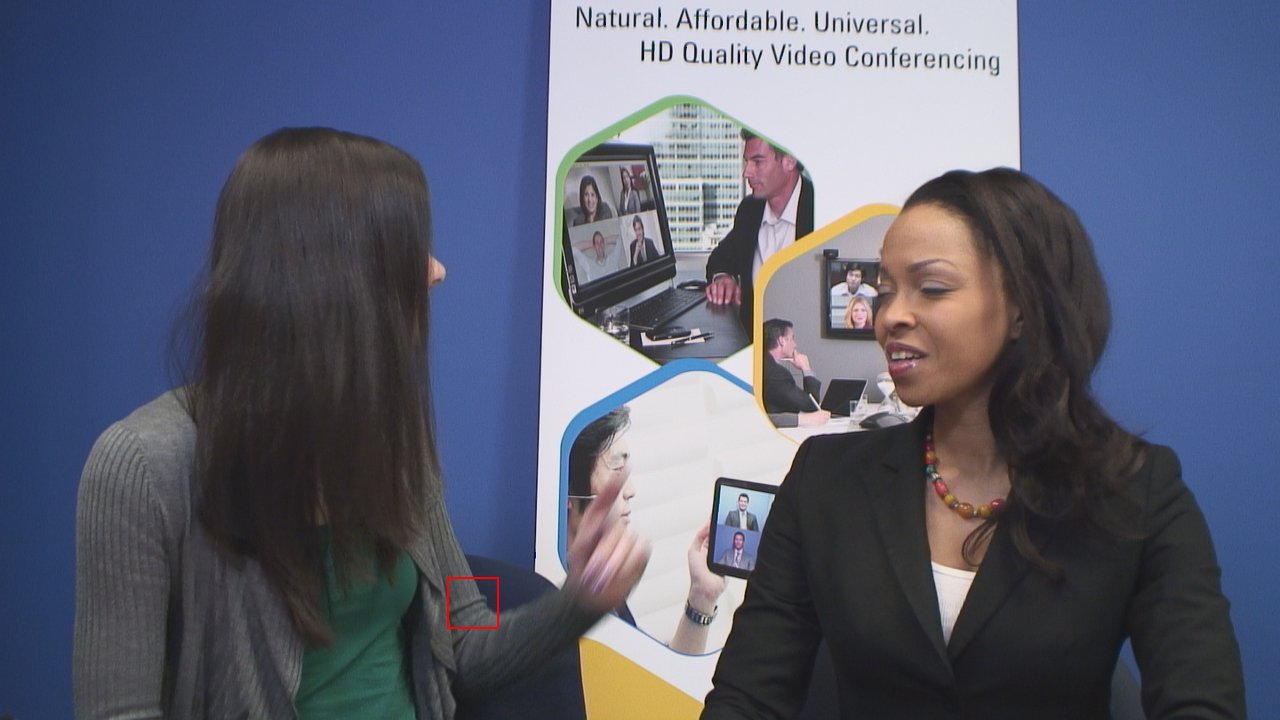,width=3.32cm}}
		\footnotesize{KristenAndSara\_142  \  }  %\medskip
	\end{minipage}
	\begin{minipage}[b]{0.11\linewidth}
		\centering
		\centerline{\epsfig{figure=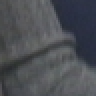,width=1.9cm}}
		\footnotesize{GT \\  } %\medskip
	\end{minipage}
	\begin{minipage}[b]{0.11\linewidth}
		\centering
		\centerline{\epsfig{figure=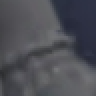,width=1.9cm}}
		\footnotesize{ Compressed  \\  }%\medskip
	\end{minipage}
	\begin{minipage}[b]{0.11\linewidth}
		\centering
		\centerline{\epsfig{figure=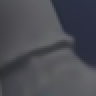,width=1.9cm}}
		\footnotesize{ STDF-R3L \\  }%\medskip
	\end{minipage}
	\begin{minipage}[b]{0.11\linewidth}
		\centering
		\centerline{\epsfig{figure=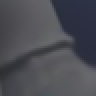,width=1.9cm}}
		\footnotesize{ RFDA \\   }%\medskip
	\end{minipage}
	\begin{minipage}[b]{0.11\linewidth}
		\centering
		\centerline{\epsfig{figure=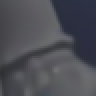,width=1.9cm}}
		\footnotesize{ CF-STIF \\  }%\medskip
	\end{minipage}
	\begin{minipage}[b]{0.11\linewidth}
		\centering
		\centerline{\epsfig{figure=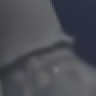,width=1.9cm}}
		\footnotesize{ STDR \\  }%\medskip
	\end{minipage}
	\begin{minipage}[b]{0.11\linewidth}
		\centering
		\centerline{\epsfig{figure=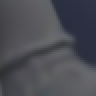,width=1.9cm}}
		\footnotesize{ Ours \\  }%\medskip
	\end{minipage}
	\caption{Visual results obtained by using different enhancement methods. The settings in the LQ sequences: BasketballPass at QP = 22 (RA), Cactus at QP = 27 (LDB), RaceHorses  QP = 32 (RA), and KristenAndSara at QP = 37 (LDB).}
	\label{RST_vision}
 \vspace{-0.5em}
\end{figure*}

\noindent
\textbf{Qualitative Results.}
We present the qualitative results of the compared methods and our method under different coding configurations and QPs in~\cref{RST_vision}. 
It is found from~\cref{RST_vision} that our method obviously reduces more compression artifacts and generates better visual results.

\vspace{-1.0em}
\subsubsection{Model Complexity}
\vspace{-0.5em}

{As described in \cite{ref24,ref25},  the evaluation of model complexity involves considering both the model parameters and the processed frames per second (FPS). The corresponding results are given in~\cref{tab_complex}.}
It is found from Tab.~\ref{tab_complex} that our method outperforms CF-STIF in terms of both parameters and FPS. The parameters of our CPGA model are similar to STDR, but CPGA achieves a higher inference speed. Overall, our CPGA demonstrates a superior balance between performance and efficiency.

\subsection{Ablation Studies}
\noindent
\textbf{Effectiveness of coding priors.}
We firstly implement a baseline model without coding priors (Model-1). Subsequently, we gradually introduce MVs and predictive frames into the IFA module of our baseline and the current residual frame into the MNA module of our baseline to verify the effectiveness of three coding priors (Model-2, Model-3, and Model-4). Besides, we incorporate two coding priors, i.e., MVs and predictive frames or predictive frames and current residual frame to construct two new models, denoting as Model-5 and Modle-6, respectively. 
Their results and corresponding model parameters are provided in~\cref{Ablation_codingprior}.  The average $\Delta$PSNR of all testing sequences under LDB configuration at QP = 37 is used to evaluate the performance gain.  It is found from~\cref{Ablation_codingprior} that our baseline achieves 0.69dB in terms of $\Delta$PSNR. Compared with STDF-R3L \cite{ref24}, our baseline saves 33K parameters and achieves better performance.  Furthermore, as can be seen from~\cref{Ablation_codingprior}, MVs, predictive frames and residual frames bring 0.03dB, 0.05dB and 0.05dB performance gain in terms of $\Delta$PSNR, compared to our baseline. Benefiting from incorporating these three coding priors, our model achieves an overall performance gain of 0.13dB in terms of $\Delta$PSNR.

\begin{table}[!t]
        \setlength{\abovecaptionskip}{0.1cm}
	\setlength{\belowcaptionskip}{0.1cm}
	\caption{The comparison of parameters and FPS.} \label{tab_complex} 
	\setlength{\tabcolsep}{2.0mm}
	\renewcommand\arraystretch{1.2}
	\fontsize{7}{9}\selectfont
	\centering
	\begin{tabular}{ccccccccc}
		\cline { 1 - 5 }
		\multirow{2}{*}{{Method}}     &\multirow{2}{*}{{Param.(K)}}   & \multicolumn{3}{c}{\text { FPS @ Different Resolution }} \\
		\cline { 3 - 5 } &    & 832$\times$480   & 416$\times$240   &   1280$\times$720  \\
		\cline { 1 - 5 }
		{ STDF-R3L } & 1275     & 11.21  & 43.53  & 4.78  \\
		{ RFDA } & 1270    & 17.91  & 59.69   & 7.08   \\
		CF-STIF  & 2200     &  4.17  & 16.32    & 1.78  \\
		STDR     & 1324     &  8.77  & 33.95  & 3.65 \\
		\cline { 1 - 5 }
		{ Ours} & 1386   & 10.42 & 39.11    & 4.08 \\
		\cline { 1 - 5 }
	\end{tabular}
        \vspace{-1.5em}
\end{table}

\begin{table}[!t]
	% \vspace{-1.5em}
	\setlength{\abovecaptionskip}{0.1cm}
	\setlength{\belowcaptionskip}{0.1cm}
	\caption{The ablation study of coding priors.} \label{Ablation_codingprior}
	\setlength{\tabcolsep}{0.4mm}
	\renewcommand\arraystretch{1.1}
	\fontsize{7}{9}\selectfont
	\centering
		\begin{tabular}{ccccc|cccc}
			\hline
			&  &  MV &  P-frame & R-frame &  { Param. (K) }   &  { $\Delta$PSNR }({dB})  &  { $\Delta$SSIM}({$\times$$10^{-2}$})\\
			\hline
			& Model-1 & - & - & -   & 1242 & 0.69  & 1.40 \\
			& Model-2 &  \checkmark & - & -  & 1348  & 0.72  & 1.42 \\
                & Model-3 &  - & \checkmark  & -  & 1348  & 0.73 & 1.44 \\
                & Model-4 &  - & -  & \checkmark  & 1280  & 0.72 & 1.42 \\
			& Model-5 &  \checkmark & \checkmark  & -  & 1348  & 0.77& 1.50\\		
			& Model-6 &  - & \checkmark  & \checkmark  & 1386  & 0.79 & 1.52 \\
			& Model-7 &  \checkmark & \checkmark & \checkmark   & 1386  & 0.82 & 1.56  \\
			\hline
		\end{tabular}
	\vspace{-1.5em}
\end{table}

\begin{figure}[!t]
        \vspace{-0.5em}
        \setlength{\abovecaptionskip}{0.1cm}
	\setlength{\belowcaptionskip}{0.1cm}
        \begin{minipage}[b]{0.19\linewidth}
		\centering
		\centerline{\epsfig{figure=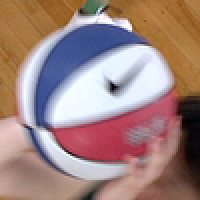,width=1.6cm}}
		\footnotesize{GT \\  } %\medskip
	\end{minipage}
	\begin{minipage}[b]{0.19\linewidth}
		\centering
		\centerline{\epsfig{figure=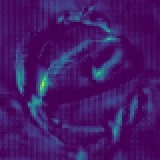,width=1.6cm}}
		\footnotesize{Model-1 \\  } %\medskip
	\end{minipage}
	\begin{minipage}[b]{0.19\linewidth}
		\centering
		\centerline{\epsfig{figure=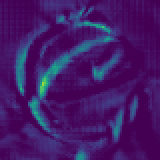,width=1.6cm}}
		\footnotesize{ Model-2  \\  }%\medskip
	\end{minipage}
	\begin{minipage}[b]{0.19\linewidth}
		\centering
		\centerline{\epsfig{figure=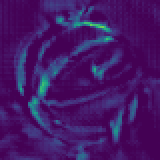,width=1.6cm}}
		\footnotesize{ Model-5 \\  }%\medskip
	\end{minipage}
	\begin{minipage}[b]{0.19\linewidth}
		\centering
		\centerline{\epsfig{figure=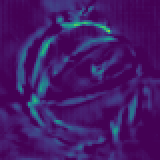,width=1.6cm}}
		\footnotesize{ Model-7 \\   }%\medskip
	\end{minipage}
	\caption{Feature visual results obtained by using coding priors.}
	\label{feat_vision}
 \vspace{-1.0em}
\end{figure}

We also provide some feature visual results using different coding priors in~\cref{feat_vision}. Specifically, we illustrate the temporally-aggregated feature $\mathcal{F}^{ta}$ of Model-1, Model-2 and Model-5, and the spatially-aggregated feature $\mathcal{F}^{sa}$ of Model-7. It is found from~\cref{feat_vision} that using MVs and predictive frames, the objects in temporally-aggregated feature $\mathcal{F}^{ta}$ have the more details. Moreover, compared with the results of Model-2 and Model-5, clear details are constructed in the feature results of Model-7, which benefits from introducing the current residual frame and our MNA module.

\noindent
\textbf{Effectiveness of VCP dataset.}
We use the LDP configuration in MFQE dataset \cite{ref23} to extract three coding priors, i.e., MV, predictive frame and residual frame, from the VCP dataset and denote this dataset as VCP(LDP) dataset. Based on this dataset, we train three state-of-the-art methods, i.e., STDF-R3L, CF-STIF, STDR, and our CPGA model at QP=37 to validate the effectiveness of our VCP dataset under LDP configuration. The corresponding results are given in~\cref{Compare}. 
In addition, we extract these three coding priors from the MFQE dataset with LDP configuration in~\cite{ref23} under QP=37 and denote this dataset as MFQE-CP to train our CPGA. The corresponding results of our CPGA are provided in~\cref{Compare}. We also offer the results of STDF-R3L~\cite{ref24}, CF-STIF~\cite{ref26}, and STDR~\cite{ref27} in~\cref{Compare} on MFQE dataset~\cite{ref23} for fair comparison, which were obtained from their respective papers. 
It is found from~\cref{Compare} that these compared methods achieve a higher performance gain by using our VCP dataset than the MFQE dataset, which demonstrates the effectiveness of our VCP dataset. It is also shown from~\cref{Compare} that our CPGA  achieves a performance gain of 0.02dB compared to the state-of-the-art method, ie., STDR, on the MFQE-CP dataset, which further verifies the effectiveness of our CPGA.

\begin{table}[t]
	\vspace{-1.0em}
	\setlength{\abovecaptionskip}{0.1cm}
	\setlength{\belowcaptionskip}{0.1cm}
	\caption{Quantitative comparison on  MFQE-CP dataset and VCP(LDP) dataset in terms of $\Delta$PSNR (dB) and $\Delta$SSIM ($\times 10^{-2}$) LDP configuration at QP=37.} \label{Compare}
	\setlength{\tabcolsep}{0.7mm}
	\renewcommand\arraystretch{1.3}
	\fontsize{7}{9}\selectfont
	\centering
	\begin{tabular}{ccccccccccccccc}
		\hline
		&  \multirow{2}{*}{\textbf{Training Dataset}}  & \multicolumn{2}{c}{\textbf { STDF-R3L }}  &\multicolumn{2}{c}{\textbf { CF-STIF }}  &\multicolumn{2}{c}{\textbf { STDR }}  &\multicolumn{2}{c}{\textbf { Ours }}  \\
		 \cline{3-4}   \cline{5-6}    \cline{7-8}  \cline{9-10} 
		&  & PSNR & SSIM  & PSNR & SSIM  & PSNR & SSIM  & PSNR & SSIM  \\
		\hline
		&  MFQE-CP & 0.83 & 1.53 & 0.92 & 1.67   & 0.98 & 1.79    & 1.00 & 1.82   \\
		\hline
		&  VCP(LDP) & 0.85 & 1.57 & 0.95 & 1.71   & 1.00 & 1.81    & 1.03 & 1.85   \\
		\hline
	\end{tabular}
 \vspace{-1.5em}
\end{table}

\vspace{-0.8em}
\section{Conclusion}
\vspace{-0.5em}
In this paper, we propose a compressed video quality enhancement dataset that includes LQ videos and their abundant coding priors.
Based on this dataset, we design a novel video quality enhancement model, named CPGA. Our proposed CPGA can effectively aggregate inter-frame temporal correlations and spatial correlations with the guidance of these coding priors to generate the HQ videos. Extensive experiments demonstrate that the CPGA achieves the state-of-the-art method for VQE. 

\noindent
\textbf{Acknowledgement.}
Thank Shijin Huang for the contributions to the dataset. This work was supported by
the National Natural Science Foundation of China under
Grant U20A20184 and the Natural Science Foundation of Sichuan Province under Grant 2023NSFSC1972.
{
   \small
   \normalem
   \bibliographystyle{ieeenat_fullname}
   \bibliography{main}
}
\end{document}